\documentclass[article,preprint,superscriptaddress]{revtex4-2}
\usepackage{hyperref}
\usepackage[english]{babel}
\usepackage{amsmath}
\usepackage{amssymb}
\usepackage{graphicx}
\usepackage{siunitx}
\usepackage{textcomp,xcolor}
\usepackage{textgreek}
\usepackage{blindtext}
\newcommand{\ignore}[1]{}

\begin{document}

\title{Impact of Spin-Orbit Coupling on Quantum Transport in Magnetic Tunnel Junction with an anti-ferromagnetic Capping Layer}

\author{Shradha Chandrashekhar Koli}
\affiliation{School of Integrated Circuit Science and Engineering, MIIT Key Laboratory of Spintronics, Beihang University, Beijing 100191, China}

\author{Hangyu Zhou}
\affiliation{School of Integrated Circuit Science and Engineering, MIIT Key Laboratory of Spintronics, Beihang University, Beijing 100191, China}

\author{Bertrand Dup\'e}é
\affiliation{Fonds de la Recherche Scientifique (FNRS), B-1000 Brussels, Belgium}
\affiliation{Nanomat/Q-mat/CESAM, Universit\'e de Li\`ege, B-4000 Sart Tilman, Belgium}
\affiliation{Institut fur Physik, Johannes Gutenberg Universit\"at Mainz, D-55099 Mainz, Germany}

\author{Weisheng Zhao}
\affiliation{School of Integrated Circuit Science and Engineering, MIIT Key Laboratory of Spintronics, Beihang University, Beijing 100191, China}

\date{\today}

\begin{abstract}
As capping layer heavy-metals with strong spin-orbit coupling (SOC) are considered essential elements in a magnetic tunnel junction (MTJ), so are the antiferromagnetic (AFM) metals in spin-orbit torque (SOT)-based devices. We explore the role of an AFM capping layer with strong SOC i.e. L1$_0$-IrMn in a perpendicularly magnetized tunnel junction (\emph{p}-MTJ), using first-principles calculations. A comparative study is conducted by employing conventional non-magnetic heavy-metals as capping layers along with IrMn in a symmetric-MTJ structure X/FeCo/MgO/FeCo/X where X=Ta, W, Mo or IrMn. Firstly, the analysis without SOC is presented where the highest magnitude of TMR is achieved in IrMn-IrMn MTJ. This large TMR originates from: (a) the large spin-polarization determined by the lattice-mismatch strain and (b) the non-identical antiparallel channels due to the antiferromagnetic ordering of IrMn that helps in maintaining lower total antiparallel conductance. Upon including SOC, the TMR increases in all the MTJ structures with - in particular - a giant enhancement in IrMn-IrMn MTJ. This relative increase in TMR is attributed to the SOC-induced (a) lifting of degeneracy which creates additional bands with \emph{p}$_z$, \emph{d}$_{z^2}$ and \emph{d}$_{xz}$ orbital characters and (b) spin- and orbital-mixing which offers additional contributors to the majority-spins channel in parallel configuration. Finally, we perform non-equilibrium transport calculations in IrMn-IrMn MTJ, which equally demonstrate an increase in TMR when SOC is switched-on. When bias voltage is applied, upon application of SOC, parallel current improves at every finite voltage while, beneficially, the antiparallel current hardly changes. Apparently, our results suggest that AFM-IrMn is a potential capping metal which can offer giant TMR in future spin-orbit toque (SOT) based MRAM devices with a straightforward design strategy.
\end{abstract}

\maketitle
\noindent

\textbf{E-mail:} \href{https://outlook.live.com/mail/}{kshradhachandrashekhar@outlook.com}


\section{Introduction}
The effect of capping layer was first discovered on tunneling magneto-resistance (TMR) by Tsumekawa et al. \cite{1464430} and on perpendicular magnetic anisotropy (PMA) by Ikeda et al. \cite{ikeda2010perpendicular} more than a decade ago. Since then, variety of heavy-metal capping layers have been tested, both, theoretically and experimentally, in search of a optimum capping material. Among these, Ta, W and Mo have been comprehensively studied as buffer-, capping-, and/or insertion layers in FeCo(B)/MgO-based devices \cite{cheng2011effect,almasi2015enhanced,lee2016perpendicular,watanabe2016magnetic}. Ta was considered a generic choice for FeCo(B)-based \emph{p}MTJs because of its good quality growth on CoFeB \cite{ikeda2010perpendicular,worledge2011spin}, while W and Mo revealed robust solution to the thermal incompatibility of Ta presenting increased PMA and TMR \cite{lee2016perpendicular,almasi2015enhanced,watanabe2017annealing}. Alternatively, due to large spin-orbit coupling (SOC) constant, W became a prominent member for spin-orbit torque (SOT)-based FeCo(B)/MgO devices \cite{pai2012spin,skowronski2015underlayer,he2016spin}.
At the same time, Ir was also found to drastically influences both the PMA and TMR in SOT-\emph{p}MTJ \cite{odkhuu2013extremely} as a result of its large SOC and the induced magnetic moments at Ir-Fe interface. The 5\emph{d}-3\emph{d} hybridization between Ir and a Ferromagnetic (FM) metal is at the origin of the enhanced PMA and TMR \cite{odkhuu2013extremely,zhou2019large}. 

Further, antiferromagnetic materials are being explored as they provide technologically potential platform for future SOT device developments \cite{guo2021spintronics}. Intriguing spin-dependent properties, high data storage density and, most importantly the capability of device downscaling make antiferromagnets more prospective over ferromagnetic counterparts. Despite advantageous device features, manipulation of magnetization switching in an antiferromagnetic heterostructure appeared as a challenge. Until recently, Peng et al. \cite{peng2020exchange} demonstrated the SOT-controlled independent switching of a FM and an AFM layer in a simple exchange-biased IrMn/CoFeB/MgO structure. As a result, switching between parallel and antiparallel resistance states in a FM layer seems realizable despite being coupled with an adjacent AFM layer.  In this work, we investigate the TMR in a symmetric IrMn/FeCo(B)/MgO/FeCo(B)/IrMn full-MTJ. We speculate that such symmetric MTJs can become useful in future SOT devices owing to their simple crystal structure with cumulative advantage of independent electrical switching of SOT and the exchange-bias presenting giant TMR \cite{zhou2018tunneling,zhou2020large}.

In this paper, we reveal the influence of an antiferromagnetic IrMn as a capping layer on the tunneling magnetoresistance. As SOC is inevitably present in real-time systems and we sought for future SOT-device structure. Very few studies \cite{popescu2005influence,yang2011first, lu2012spin} have reported the effect of SOC in capping layers on spin transport in Fe-based structures. When SOC is switched-on, the understanding of tunneling with spin-mixed waves becomes complex. Therefore, we investigate both the spin-decomposed transport and the SOC-induced spin-mixed transport in X/FeCo/MgO/FeCo/X MTJ employing capping layers such as X=Ta, W, Mo and IrMn with different degree of SOC strength.  This paper is decomposed as follows: First Section describes the methodology used to construct energetically stable MTJ-stacks and to compute spin-dependent transport properties, in second section we discuss obtained results of tunneling before and after including SOC factor.

\section{Methods}

\subsection{DFT calculation}
The relaxed electronic structures were obtained with the projector augmented wave (PAW)\cite{kresse1996efficiency} Pseudopotentials as implemented in Vienna Ab Initio Simulation Package (VASP)\cite{hafner2008ab}. The exchange-correlation energy was treated within the generalized gradient approximation (GGA) using Perdew-Burke-Ernzerhof (PBE) functionals \cite{perdew1992accurate,perdew1996generalized}. The convergence threshold for total atomic energies and Hellmann-Feynman forces are set to 10$^{-6}$ eV and 0.01 eV.{{\AA}}$^{-1}$, respectively. Kohn-Sham orbitals were expanded using the plane-wave basis set with kinetic energy cutoff of 550 eV. The sampling of the Brillouin zone is obtained with a $\Gamma$-centered Monkhorst-Pack scheme \cite{monkhorst1976special} of $25 \times 25 \times 1$ k-points. This ensures sufficient convergence of the forces. The illustration of the atomic structures was obtained using VESTA \cite{momma2011vesta} software.

\subsubsection{Magnetic Stability of B$_{2}$-IrMn}

\begin{figure}[ht]
\centering
\includegraphics[width = 0.7\columnwidth]{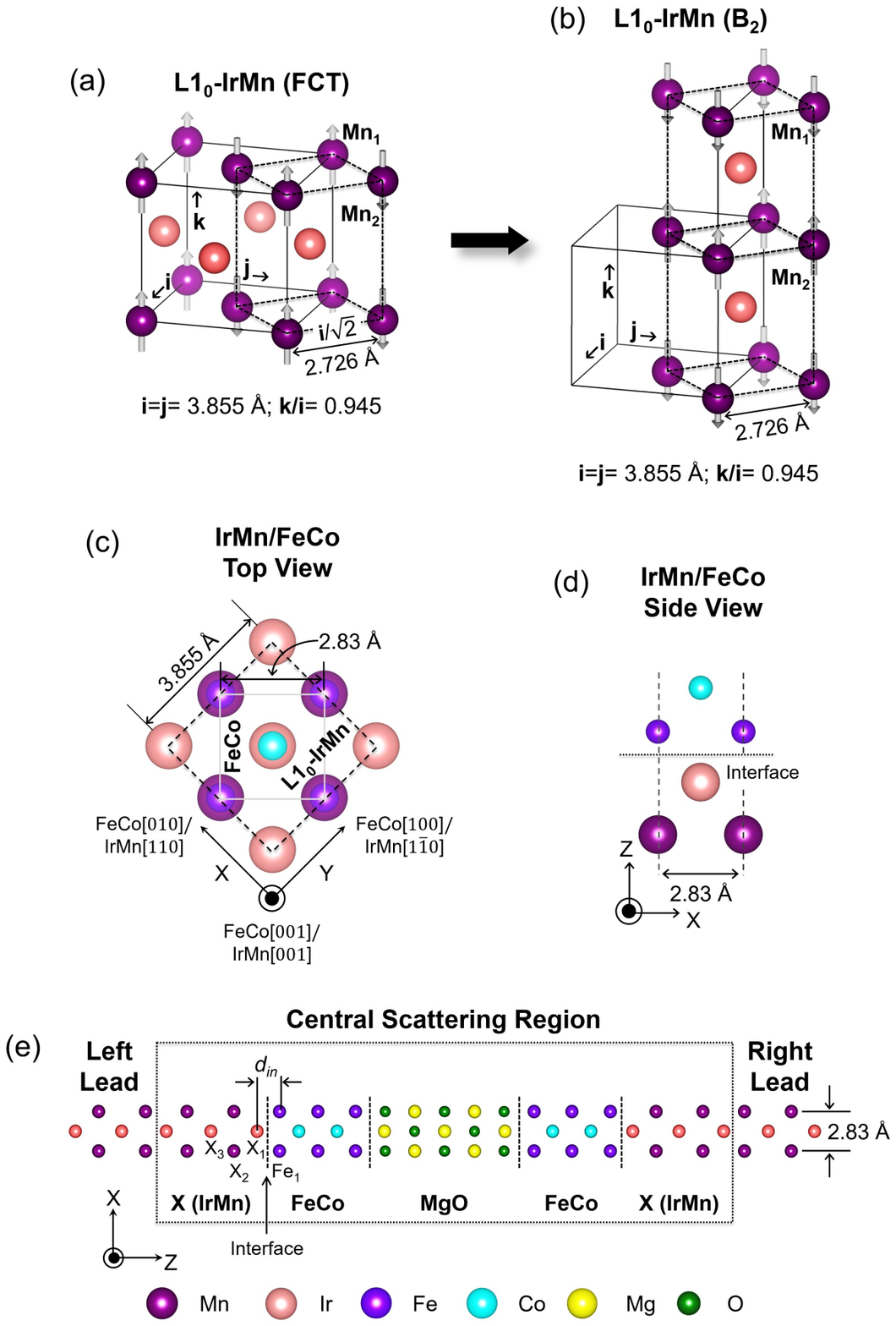}
\caption{Illustration of the energetically favored  crystallographic structures. (a) L1${_0}$-IrMn FCT cell with anti-ferromagnetically  aligned Mn$_1$ (up) and Mn$_2$ (down) spins. (b) A magnetically stable B${_2}$ unit cell implemented in our calculation. (c) An xy-plane crystallographic relationship of IrMn($001$)/FeCo($001$) with (d) Ir-Fe terminated interfacial arrangement along the direction of transport. The direction of transport is along the z-axis. (e) A two-probe model of CoFe/MgO/CoFe MTJ-stack with 'X' capping layer (here IrMn).  A collinear magnetic spin-alignment normal to the interface is considered. X${_1}$, X${_2}$, and X${_3}$ represent first three atomic positions within the capping layer starting from the FeCo interface. $d_{\mathrm{in}}$ denotes the interface atomic distance (X${_1}$-Fe${_1}$) between X/FeCo with Fe${_1}$ being an interface terminating atom.} 
\label{fig:structure}
\end{figure}

In order to maintain a relatively small lattice-mismatch with the FeCo/MgO/FeCo junction, the IrMn L1${_0}$ structure in Fig.\ref{fig:structure}(a) is rotated 45$^{\circ}$ in-plane \cite{zhang2017first,han2018effects}, to obtain a B${_2}$ structure with lattice constant $\bold{i} / \sqrt{2}$=2.73 {\AA}. The electronic structural details of L1$_0$-IrMn are presented in Appendix A. Unlike typical L1$_0$-IrMn, the ($001$)[$110$] oriented B${_2}$ unit cell consists of alternatively arranged mono-atomic Ir- and Mn- planes. This alters the spin-alignment of Mn-atoms along the ($001$) direction \cite{pal1968magnetic,wang2013structural,wang2003ab}. Fig.\ref{fig:structure}(b) shows the magnetic unit cell of B${_2}$-IrMn with AFM spin-alignment of Mn-atoms.  Because of the alternating arrangement of Ir- and Mn- planes, a full collinear magnetic relaxation was performed in two different sequences of 5ML-stack with alternating Ir- and Mn terminations, to confirm the AFM ordering. The first sequence with 2ML-Ir and 3ML-Mn produced a finite net magnetic moment making 5ML-IrMn capping a ferromagnet. Whereas, the sequence with 3ML-Ir and 2ML-Mn resulted in a 0.0 $\mu_B$ net magnetic moment implying that a pure Ir-terminated 5ML-stack is more stable over the Mn-terminated one as it retains the antiferromagnetic nature of B${_2}$-IrMn capping. A large spin magnetic moment of $\sim$2.8 $\mu_B$ on Mn-atoms is obtained after relaxation \cite{szunyogh2009giant}. In addition, the lattice-mismatch is minimized to -2.46\% which is smaller compared to FeCo/MgO (+4.7 \%) or Ta/FeCo, W/FeCo, Mo/FeCo structures.

\subsubsection{MTJ stacking}
As in real systems, a heavy-metal deposited below (seed layer) and above (capping layer) a FM-electrode have different influence \cite{cheng2011effect,liu2018effect,zhang2019effects}, a full MTJ supercell was structurally relaxed until the forces acting on each atom were smaller than 0.01 eV.{\AA}$^{-1}$. Here we refer to both the seed and the capping layer as capping layers throughout the paper as we consider symmetric MTJs X(5)/FeCo(5)/MgO(5)/FeCo(5)/X(5) where X= Ta, W, Mo and IrMn. The numbers in the parentheses represent the number of monolayers employed for each layer in the MTJ structure. The lattice constant of the non-magnetic capping layers, Ta, W and Mo, are 3.317 {\AA}, 3.187 {\AA} and 3.168 {\AA}, respectively. Also, a 15 {\AA} of vacuum layer is included on top of all the stacks to ensure no interactions between the different images along the z-direction. Hereafter, the respective full-MTJ stacks will be referred to as simply Ta-Ta, W-W, Mo-Mo and IrMn-IrMn MTJ.

Table \ref{table:II} summarizes the longitudinal interatomic distances along the (001)-direction. In the relaxed MTJ, the interlayer distance ($d_{\mathrm{cap}}$) within the non-magnetic capping layers Ta, W and Mo has remarkably increased  to that in the bulk ($d_{\mathrm{b}}$) which can be seen as the after-effect of the in-plane compression of the lattice due to a huge lattice-mismatch. Meanwhile, the slightly reduced $d_{\mathrm{cap}}$ within the IrMn capping can be associated to an equal and slight expansion of in-plane lattice which causes a longitudinal shrinkage. Concurrently, the interfacial Ir-Fe distance, despite with an ideal lattice-mismatch, seems to have relatively increased. This increase can be attributed to a trade-off between the Ir-Fe bonding on one side and the Ir-Mn on the other which, further, can be inferred from the direct correspondence between the $d_{\mathrm{in}}$ around X/FeCo interface and the ($d_{\mathrm{b}}$) in the bulk. 

Furthermore, the energetic stability of IrMn/FeCo and FeCo/MgO interfaces is also inspected. in the IrMn-incorporated bilayer and trilayer structures. See Appendix B.

\begin{table}[h!]
\caption{The out-of-plane inter-atomic distance along the direction of transport. The $d_{\mathrm{b}}$ is an interlayer distance within the bulk capping layer. The inter-atomic distance ($d_{\mathrm{cap}}$) within the capping layer and the interface distance ($d_{\mathrm{in}}$) between the capping and FeCo layer in the relaxed MTJ structures as shown in Fig.\ref{fig:structure}(e).}
\centering
\begin{tabular}{m{1.1cm} m{1.0cm} m{2cm} m{1.0cm} m{1.0cm} m{1.0cm}} 
 \hline\hline \\ 
 Bulk & $d_{\mathrm{b}}$ (\AA) & MTJ & \multicolumn{2}{c}{$d_{\mathrm{cap}}$ (\AA)} & $d_{\mathrm{in}}$ (\AA) \\
 \hline \\ 
{} & {} & {} & X${_3}$-X${_2}$ & X${_2}$-X${_1}$ & X${_1}$-Fe${_1}$\\
 \hline \\ 
 Ta-Ta & 1.66 & Ta-Ta & 2.35 & 2.32 & 1.86  \\
 W-W & 1.59 & W-W & 2.10 & 2.09 & 1.62 \\ [0.5ex]
 Mo-Mo & 1.58 & Mo-Mo & 2.05 & 2.09 & 1.62 \\ [0.5ex]
 Ir-Mn & 1.82 & IrMn-IrMn & 1.72 & 1.73 & 1.67 \\ [0.5ex]
 \hline\hline
 \label{table:II}
\end{tabular}
\end{table}

\subsection{Transport calculations}

The quantum transport properties of electronically stable MTJs are calculated via the Keldysh-nonequilibrium-Green’s-function formalism combined with real-space density functional theory (NEGF-DFT) as implemented in Nanodcal\cite{taylor2001ab,taylor2001ab} package. The device model in Nanodcal is divided into three sections: the central scattering (CS) region and the two leads (capping layers).  The CS region includes 5ML of the capping layer as shown in Fig.\ref{fig:structure}(e). Self-consistent calculations were performed using the double plus polarized (DZP) atomic orbitals basis with electron kinetic energy cut-off of 5000 eV. The same PBE-GGA functional was used in both structural relaxation and the quantum transport calculations. SOC was included self-consistently.

The spin-polarized conductance G$_{\sigma}$ is given by the Landauer-B\"uttiker formula as,

\begin{eqnarray}
G_\sigma = \frac{e^2}{h} \sum_{k_\parallel} T_\sigma (k_\parallel , E_F) ,
\label{eq:1}
\end{eqnarray}

where $T_{\sigma} (k_{\parallel} , E_F)$ is the transmission coefficient at Fermi Energy $E_F$  with spin $\sigma=(\uparrow,\downarrow)$ and an in-plane Bloch wave vector $k_{\parallel}=(k_x,k_y)$. $e$ is the charge of an electron and $h$ the Plank’s constant. During self-consistent calculations, the Hamiltonian matrix and the density matrix were converged down to less than 10$^{-4}$ Hartree with a $k_{\parallel}$-mesh sampling of $10 \times 10 \times 1$ for CS region and $10 \times 10 \times 100$ for the leads. The transmission coefficient and conductance were evaluated using denser $k_{\parallel}$-mesh of $300 \times 300 \times 1$ for all spin-channels, without and with SOC. 

The tunneling magnetoresistance (TMR) is defined as,

\begin{eqnarray}
TMR = \frac{G_P-G_{AP}}{G_{AP}},
\label{eq:2}
\end{eqnarray}

where $G_P=G_P^{\uparrow \uparrow}+G_P^{\downarrow \downarrow}$  is the conductance in the parallel (P) configuration of spins in FeCo electrodes. $G_P^{\uparrow \uparrow}$($G_P^{\downarrow \downarrow}$) defines the conductance when a spin travels from a majority(minority)-to-majority(minority) spins channel. Similarly, $G_{AP}=G_{AP}^{\uparrow \downarrow}+G_{AP}^{\downarrow \uparrow}$ is the conductance in the anti-parallel (AP) configuration with $G_{AP}^{\uparrow \downarrow}$($G_{AP}^{\downarrow \uparrow}$) being the conductance of  majority(minority)-to-minority(majority) channel. This implies that the effective spin-polarization (SP) is defined as

\begin{eqnarray}
SP = \frac{G_P^{\uparrow\uparrow}-G_P^{\downarrow\downarrow}}{G_P^{\uparrow\uparrow}+G_P^{\downarrow\downarrow}} .
\label{eq:3}
\end{eqnarray}
Here, the magnetic moments in the MTJs are collinear.

Unlike FeCo, Ta, W or Mo, IrMn has a tetragonal lattice (see Appendix A) with symmetry in \emph{P4/mmm} (No. 123) space group.  In this case, the orbital decomposition of the Bloch wave symmetries like $\Delta_1$ states with \emph{s},$\emph{p}{_z}$,$\emph{d}{_{z^2}}$ will not be similar to those of a square lattice as in Ref.\cite{butler2001spin}. In fact, the wave symmetries may no longer be classified as the $\Delta_1$, $\Delta_5 $, $\Delta_{2}’$ or $\Delta_2$ states, due to the possibility of mixing of other electronic orbitals. In this paper, we discuss transport by means of electronic orbitals instead of Bloch symmetries in all X-X MTJs, for the sake of comparison. One must note that, the Bloch wave symmetries for bcc Ta, W or Mo exclusively remain the same as for FeCo square lattice. We restrict our discussion to the out-of-plane oriented orbital characters like \emph{s},$\emph{p}{_z}$,$\emph{d}{_{z^2}}$,$\emph{d}{_{xz}}$,$\emph{d}{_{yz}}$ . Additionally, only the out-of-plane \emph{d}-orbitals in the X capping layer are of exclusive importance as only the 3/4/5\emph{d}-X and the 3\emph{d}-FeCo orbital hybridization influences PMA or TMR \cite{odkhuu2013extremely,peng2015origin,chen2015underlayer}. Furthermore, in-plane oriented $\emph{d}{_{xy}}$ and $\emph{d}{_{x^2-y^2}}$, orbitals which are not interacting with the out-of-plane $\emph{p}{_z}$ character in the MgO-barrier result in a rapid decay of the wave function within the barrier hindering the propagation of the tunneling wave \cite{butler2001spin}.

\section{Results and Discussion}

\subsection{Spin- and SOC-Dependent Conductance}

\subsubsection{Conductance without SOC}

The TMR in Eq.\ref{eq:2} is the ratio of conductance of two different channels with respective spin-alignments. Upon switching on SOC, explicit understanding of the spin-dependent tunneling conductance and, therefore, the TMR  becomes challenging. Accordingly, we investigate the SOC and discuss its contribution in comparison with the No-SOC results in X-X MTJs. Fig.\ref{fig:Fig2}(a) depicts the spin-dependent tunneling conductance, without SOC, for a parallel (P) and an anti-parallel (AP) spin configuration. In parallel configuration, the conductance of majority-spin channel  (G$_P^{\uparrow\uparrow}$) in the Ta-Ta MTJ is lower than in W-W, Mo-Mo or IrMn-IrMn MTJ. In fact, it is even lower than its minority-spin channel conductance (G$_P^{\downarrow\downarrow}$). The large lattice-mismatch in Ta-Ta exerts an in-plane compressive strain on the Ta lattice and pushes away the Ta atoms in z-direction,  further increasing the interface Ta-Fe distance \cite{ong2015strain} as in Table \ref{table:II}. This leads to a formation of discontinued interface between Ta and FeCo. As a consequence, the normal incidence contribution from the out-of-plane orbitals like \emph{s},$\emph{p}{_z}$,$\emph{d}{_{z^2}}$ ($\Delta_1$ states for Ta-Ta)  at $\Gamma$ ($k_\parallel=(0,0)$) in majority-spin channel decreases resulting into a larger minority-spin channel conductance \cite{sankaran2016oscillatory}. In comparison, the lattice-mismatch and the correlated interface distance  of W/FeCo is considerably smaller in comparison with the Mo/FeCo interface, owing to the almost similar lattice constant of Mo as that of W.  This is reflected as a relative increase in the magnitude of G$_P^{\uparrow\uparrow}$ in W-W and Mo-Mo MTJ. Also, a slight increase of G$_P^{\downarrow\downarrow}$ in Mo-Mo MTJ, however, breaks the trend of linearly decreasing G$_P^{\downarrow\downarrow}$.  Moreover, in IrMn-IrMn MTJ, stronger interfacial coupling due to reduced lattice mismatch should facilitate a pathway for efficient tunneling at normal incidence ($\Gamma$-point). This, in turn, should increase the magnitude of G$_P^{\uparrow\uparrow}$ \cite{sankaran2016oscillatory}. On the contrary, G$_P^{\uparrow\uparrow}$ appears to have slightly dropped as that of Mo-Mo MTJ. More importantly, unlike in other MTJs, the antiparallel conductances G$_{AP}^{\uparrow\downarrow}$ and  G$_{AP}^{\downarrow\uparrow}$ show a non-identical behavior signifying different characteristics of the majority- and minority-spins channels.  To realize an anti-parallel configuration, the spin-alignment of, only, one of the FeCo-electrode (free-layer) is flipped by $180^\circ$,  keeping the spin-alignment in IrMn capping layers unchanged.

\subsubsection{Lattice-mismatch determined Spin-Polarization}

To explain the atypical characteristics of G$_P^{\uparrow\uparrow}$/G$_P^{\downarrow\downarrow}$ in Ta-Ta, we calculate the spin-polarization (SP) in parallel configuration expressed by Eq.\ref{eq:3} for the full-MTJs  \cite{meservey1994spin}. Besides, the TMR can also be quantitatively understood by analyzing the SP of the tunneling electrons \cite{julliere1975tunneling}. It is seen in Fig.\ref{fig:Fig2}(b) that the negative SP in Ta-Ta justifies  G$_P^{\downarrow\downarrow}$ being larger than that of G$_P^{\uparrow\uparrow}$ in Fig.\ref{fig:Fig2}(a). The cause of this negative SP, and therefore, the greater G$_P^{\downarrow\downarrow}$, however, remains unknown. Although, SP is affected by many different factors, here, we attribute the cause to the lattice under strain \cite{loong2014strain}.  Our attribution to the lattice-mismatch-induced strain, originating from the capping layers, is closely related to the obtained increase in the magnitude of SP. Due to very large lattice-mismatch, the in-plane lattice of the capping layers Ta, W and Mo gets compressed to reach the fixed in-plane lattice constant of 2.83 {\AA} which reduces the in-plane interatomic distance. While the atomic distance shrinks along x-direction the atoms are forced to move in the z-direction elongating the lattice, and therefore, the interatomic distance in the z-direction increases. This implies that the compressive strain in x-direction now acts as a tensile strain in the z-direction. As a result, an increase in the interfacial X${_1}$-Fe${_1}$ distance between X/FeCo is seen in Table \ref{table:II}. As, in our study, the electrons are travelling in z-direction, the strain acting in the z-direction shows stronger impact on SP.  Cai, Y. et al.  \cite{cai2012effect} demonstrated that SP responds differently to the compressive and the tensile strain. In Ta-Ta, due to large lattice-mismatch, the tensile strain acting on the capping lattice and Ta/FeCo interface is comparatively larger than in W-W, Mo-Mo or IrMn-IrMn MTJ. Impact of this tensile strain on the Ta/FeCo interface is seen in Table \ref{table:II} where interface X${_1}$-Fe${_1}$ distance in Ta-Ta MTJ is the longest. The tensile strain, to which SP is inversely proportional, increases the interfacial states causing a reduction in the SP \cite{cai2012effect}. The contribution from interfacial states in Ta-Ta is increased, so much so that SP now becomes negative. As the contribution of these interfacial states is predominant in minority-spins channel \cite{butler2008tunneling}, the conductance G$_P^{\downarrow\downarrow}$ supersedes G$_P^{\uparrow\uparrow}$ in Ta-Ta MTJ as seen in Fig.\ref{fig:Fig2}(a). Similarly, an increase in the magnitude of SP in W-W, Mo-Mo and IrMn-IrMn MTJs is the result of corresponding decrease in the lattice-mismatch. Contrary to other MTJs, in IrMn-IrMn, the strain acting in the direction of transport is the compressive strain as the in-plane lattice constant of IrMn (2.73 {\AA}) is smaller than the fixed lattice constant (2.83 {\AA}). This compressive strain preserves stronger Ir-Fe coupling at the IrMn/FeCo interface resulting into a highest magnitude of SP.

\subsubsection{Conductance with SOC}

Next, in Fig.\ref{fig:Fig2}(c), we present the analysis of conductance upon including spin-orbit coupling (SOC). The parallel conductance G$_P$ (G$_P^{\uparrow \uparrow}$+G$_P^{\downarrow \downarrow}$) and the anti-parallel conductance G$_{AP}$ (G$_{AP}^{\uparrow \downarrow}$+G$_{AP}^{\downarrow \uparrow}$) indicate a cumulative contribution of SOC-induced spin-mixed conduction channels \cite{yang2011first}. This means that the majority(minority)-to-majority(minority) channel in parallel configuration and majority(minority)-to-minority(majority) channel in the anti-parallel configuration are indistinguishable. Primarily, G$_P$ is seen to increase in all the MTJ structures except in Mo-Mo. This increase highlights the importance of using capping metals with strong SOC constant. As one can also see, the magnitude of G$_P$ in Mo-Mo is nearly equal to the magnitude of G$_P^{\uparrow \uparrow}$ in Fig.\ref{fig:Fig2} (a) which signifies the almost negligible impact of weak SOC in Mo and hence, the absence of spin-mixed contribution in Mo-Mo MTJ on the conductance. Additionally, the graph of G$_{AP}$, after including SOC, behaves similar to that in Fig.\ref{fig:Fig2}(a) except that the non-identical signatures of G$_{AP}^{\uparrow \downarrow}$ and G$_{AP}^{\downarrow \uparrow}$ cannot be distinguished. Important to note, SOC, weak or strong, is inevitably present in real systems.

\subsubsection{Effect of SOC on TMR}

\begin{figure}[h!]
\centering
\includegraphics[width = 0.9\columnwidth]{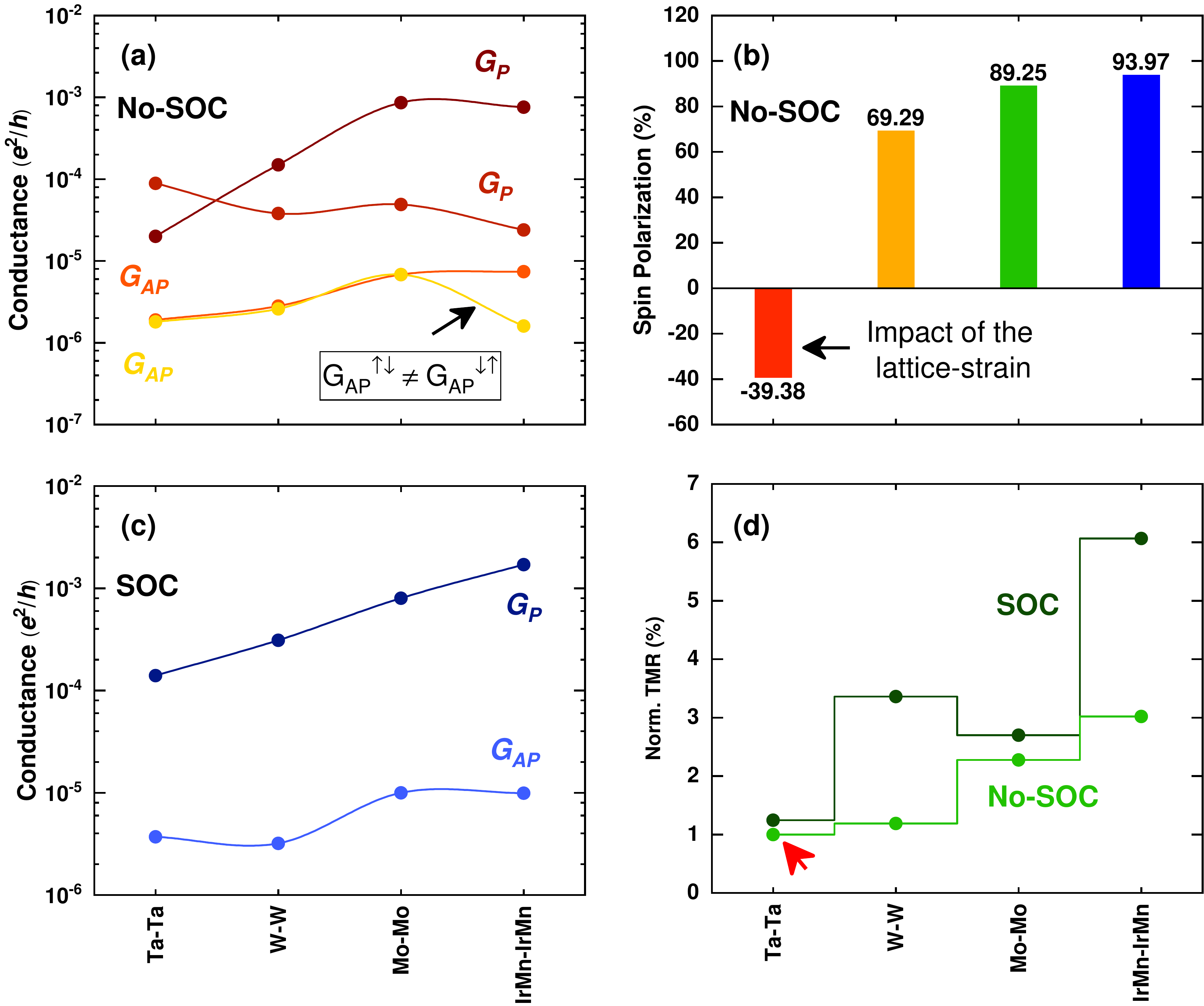}
\caption{Spin-dependent transport properties in FeCo/MgO/FeCo MTJ with different capping layers. Plots of (a) conductance in parallel (P) and anti-parallel (AP) spin configuration without SOC.(b) Spin-polarization of MTJ-stacks without SOC, using Eq.\ref{eq:3}. (c) SOC-induced conductance in P and AP spin configuration. The values of conductance are in the units of $e^2h^{-1}$. Conductance with SOC is not spin-decomposed. In (d) The Normalized TMR ratio as a function of SOC. For the sake of comparison, TMR ratios are normalized by No-SOC TMR for Ta-Ta MTJ.} The solid lines are a guide to the eye, only.
\label{fig:Fig2}
\end{figure}

In Fig.\ref{fig:Fig2}(d) the normalized TMR without and with spin-orbit coupling (SOC) is illustrated. For the sake of comparison, the TMR ratios without SOC (No-SOC) and with SOC are normalized by the No-SOC TMR obtained for Ta-Ta MTJ.  The graph of  No-SOC TMR exhibits an ascending order similar to the lattice-strain-dependent SP in Fig.\ref{fig:Fig2}(b). Although, in IrMn-IrMn, G$_P^{\uparrow \uparrow}$ and G$_P^{\downarrow\downarrow}$ show a decrease over Mo-Mo MTJ (see Fig.\ref{fig:Fig2}(a)), the No-SOC TMR, conversely, shows an increase. This increase can be ascribed to the non-identical G$_{AP}^{\uparrow \downarrow}$ and G$_{AP}^{\downarrow \uparrow}$ in IrMn-IrMn MTJ. The magnitude of G$_{AP}^{\downarrow \uparrow}$ achieved is almost, comparably, the least. Typically, the lower the antiparallel conductance (G$_{AP}^{\uparrow \downarrow}$ + G$_{AP}^{\downarrow \uparrow}$) the larger the TMR ratio obtained \cite{butler2008tunneling}. In consequence, the lowest G$_{AP}^{\downarrow \uparrow}$ and hence, the lowest G$_{AP}^{\uparrow \downarrow}$ + G$_{AP}^{\downarrow \uparrow}$ in IrMn-IrMn should simply justify the largest TMR achieved. When SOC is included, firstly, an increase in TMR ratio in all the MTJs is observed. One can attribute this increase to the SOC-induced: (a) lifting of the degeneracy which increases the band levels, especially, of the out-of-plane orbital characters \cite{yang2011first} and (b) spin- and orbital-mixing which results in mixing of Bloch wave symmetries \cite{popescu2005influence,lu2012spin}. Further, we point out that, while the increase of TMR in W-W and IrMn-IrMn MTJs is relatively significant, the difference of increase is $\sim$25\% greater in IrMn-IrMn than in W-W despite W being a comparatively stronger SOC metal. So, here, we suspect that this striking difference is not only induced by the strength of spin-orbit coupling but also, together, by the strength of interfacial coupling due to lattice-mismatch strain in the capping layers. Nevertheless, the discussion of Fig.\ref{fig:Fig2} raises questions regarding No-SOC transport, such as: (i) the cause of non-identical signatures of G$_{AP}^{\uparrow \downarrow}$ and G$_{AP}^{\downarrow \uparrow}$ in IrMn-IrMn and (ii) the enhancement of No-SOC TMR in IrMn-IrMn over Mo-Mo MTJ despite the relative decrease in $G_P^{\uparrow \uparrow}$ and G$_P^{\downarrow\downarrow}$.
 
\subsection{Energy-Dependent Transmission Coefficient}
\subsubsection{Antiparallel Transmission}
\begin{figure}[h!]
\centering
\includegraphics[width = 0.7\columnwidth]{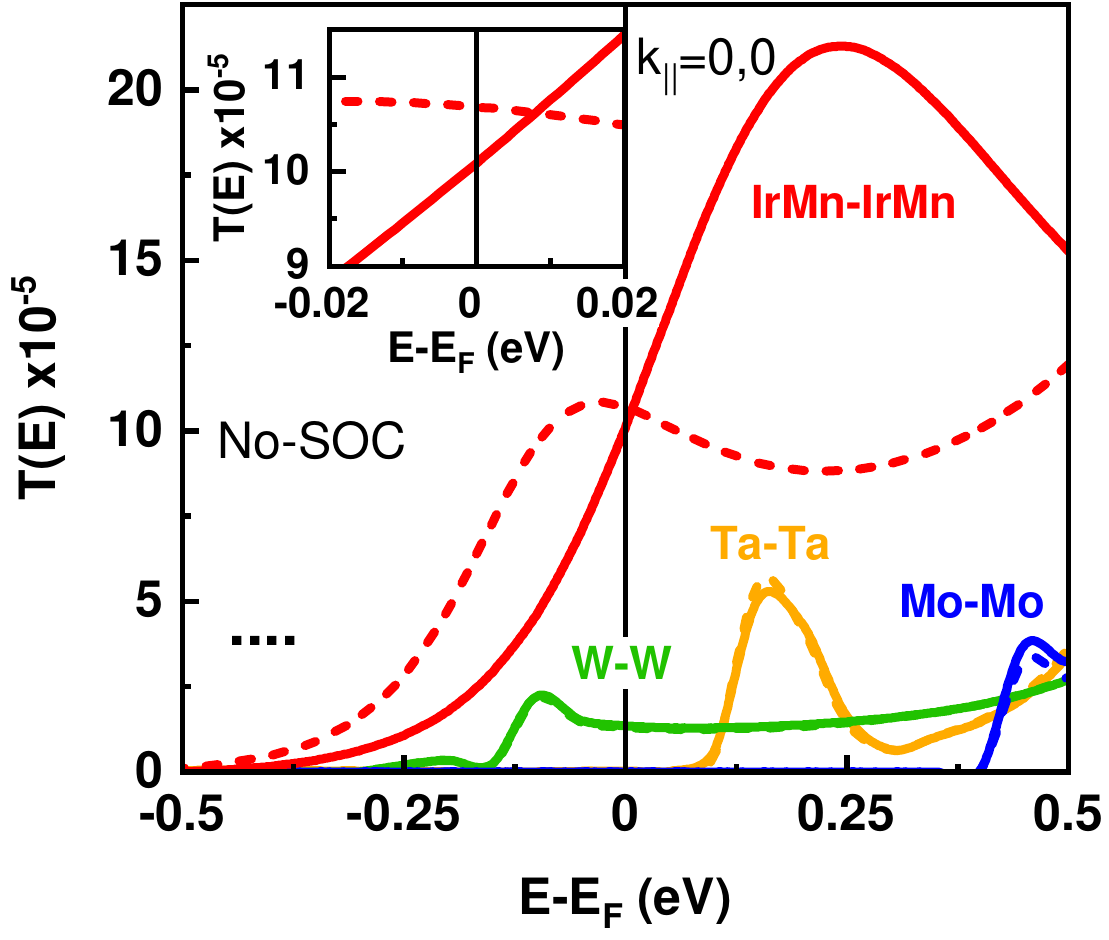}
\caption{Transmission as a function of electron energy for antiparallel channels, $\uparrow \downarrow$ and $\downarrow \uparrow$, under equilibrium. Inset: distinct crossing of $\uparrow \downarrow$ and $\downarrow \uparrow$ channels in IrMn-IrMn MTJ.  The calculations are performed specifically at $\Gamma$-point}
\label{fig:Fig3}
\end{figure}

To understand this unusual behaviour of transport, next we investigate is the transmission coefficient as a function of electron energy, T(E), without SOC. Since, the transmission at $\Gamma$-point ($k_\parallel=(0,0)$) contributes maximum to the tunneling (current), we inspect the transmission coefficient, specifically, at $\Gamma$-point. Firstly, to address the anomaly in G$_{AP}^{\uparrow \downarrow}$ and G$_{AP}^{\downarrow \uparrow}$ in Fig.\ref{fig:Fig2}(a), we analyze T(E) of AP$^{\uparrow \downarrow/\downarrow \uparrow}$ channels for X-X MTJs. Clearly, in Fig.\ref{fig:Fig3}, contrasting to Ta-Ta, W-W and Mo-Mo, the AP$^{\uparrow \downarrow}$ and AP$^{\downarrow \uparrow}$ channels in IrMn-IrMn portray a sharply distinct behavior, especially, at the Fermi energy crossing (see inset of Fig.\ref{fig:Fig3}). It is very likely that the non-identical behavior of the AP channels roots from: (a) independent contribution of the two different metallic species, Ir and Mn and/or (b) the antiferromagnetic property of IrMn capping which creates an imbalanced AFM/FM interface. These arguments can be further justified by examining the local density of out-of-plane $\emph{d}$-orbital states of X-capping layers. Corresponding local density of states (LDOS) at $\Gamma$-point projected onto $\emph{d}{_{z^2}}$ and the $\emph{d}{_{xz,yz}}$ orbital characters is illustrated in Fig.\ref{fig:Fig11}(a) and (b), respectively (see Appendix C).

\subsubsection{Majority-Spins Channel Transmission}
\begin{figure}[h!]
\centering
\includegraphics[width = 0.7\columnwidth]{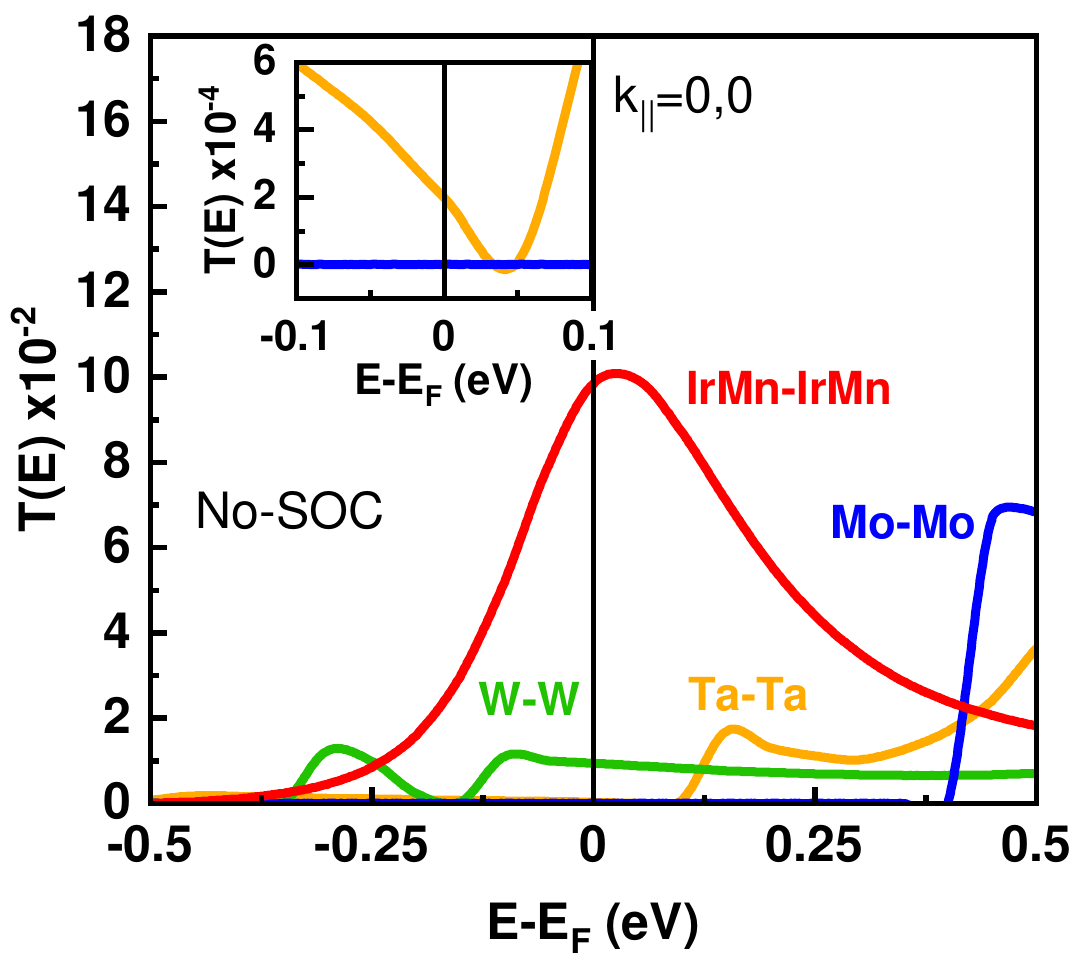}
\caption{The transmission coefficient as a function of electron energy for majority-spins channel in parallel configuration (P$^{\uparrow\uparrow}$), under equilibrium. Inset: magnitude of transmission coefficient for Ta-Ta and Mo-Mo, magnified in the area near to Fermi energy. The calculations are performed specifically at $\Gamma$-point}
\label{fig:Fig4}
\end{figure}

Second, in order to understand the increase of TMR in IrMn-IrMn as that of Mo-Mo MTJ in Fig.\ref{fig:Fig2}(d), the transmission coefficient of majority-spin channel in the parallel configuration is studied. The T(E) of P$^{\uparrow \uparrow}$ channel is represented in Fig.\ref{fig:Fig4} which is specifically calculated at $\Gamma$-point ($k_\parallel=(0,0)$). As one can see, unlike other MTJs, the transmission in IrMn-IrMn peaks very close to the Fermi energy with the largest magnitude, in comparison. The smallest magnitude of transmission is offered by Ta-Ta whereas an almost zero magnitude is displayed by Mo-Mo as shown in the inset of Fig.\ref{fig:Fig4}. As a result, the tunneling (current) which is majorly contributed by the transmission at $\Gamma$-point, is the largest in IrMn-IrMn MTJ than in Mo-Mo and others. The TMR ratio in IrMn-IrMn, hence, is the largest in Fig.\ref{fig:Fig2}(d) despite the lower magnitudes of conductance, G$_P^{\uparrow \uparrow}$ and G$_P^{\downarrow\downarrow}$. Further, the deeper understanding of transmission at $\Gamma$-point can be achieved by analyzing the local density of states (LDOS) at $\Gamma$-point. As the out-of-plane 3/4/5$\emph{d}$-X orbitals hybridize with adjacent 3$\emph{d}$-FeCo and decide the nature of transmission at $\Gamma$-point, we project LDOS onto $\emph{d}{_{z^2}}$ and the $\emph{d}{_{xz,yz}}$ orbital characters, only. Corresponding LDOS of X-capping layers is illustrated in Fig.\ref{fig:Fig12}(a) and (b) of Appendix D.

\subsection{Transmission Probability}

\subsubsection{MTJs with non-magnetic capping layers}

\begin{figure}[h!]
\centering
\includegraphics[width = 1.0\columnwidth]{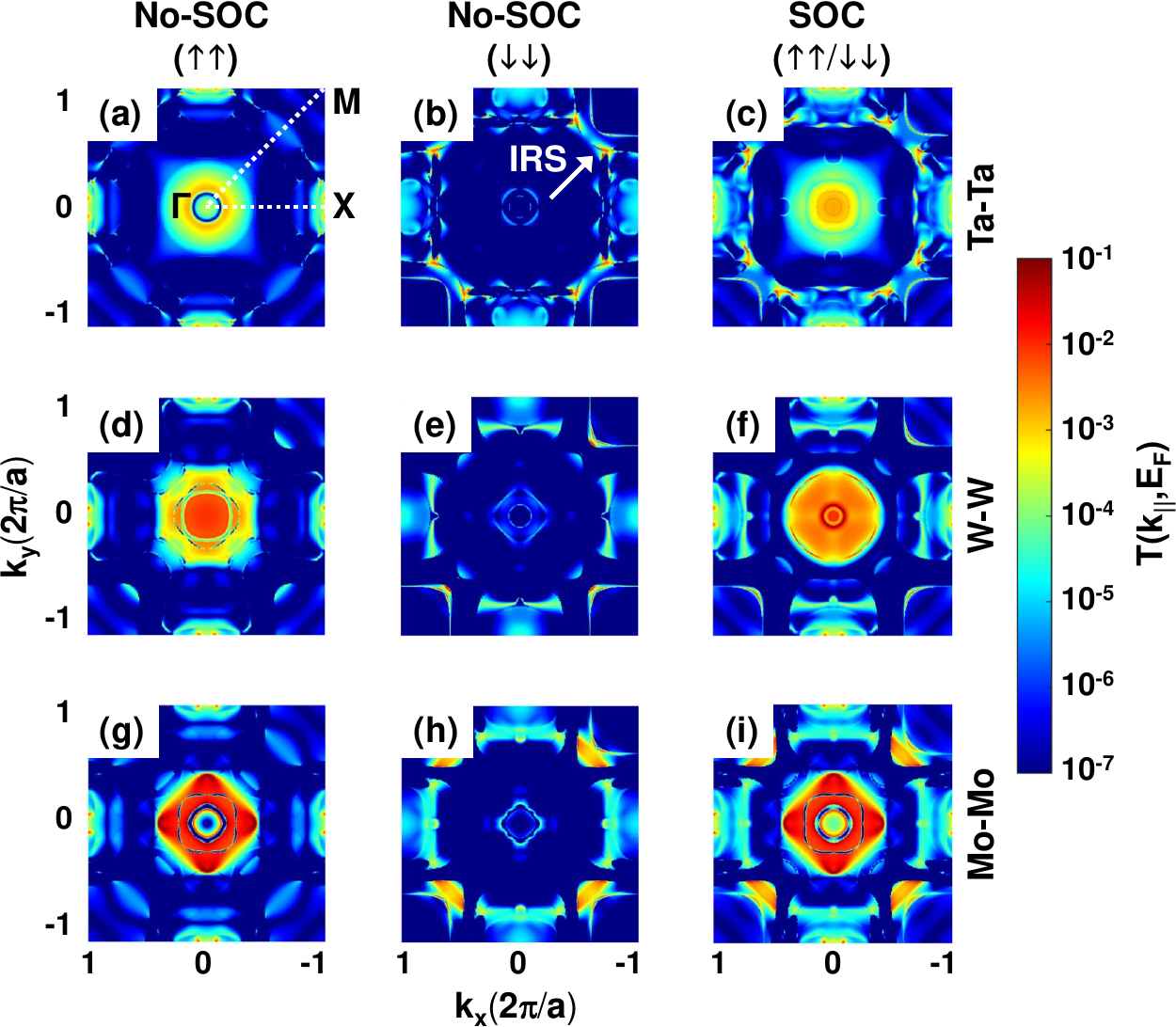}
\caption{Transmission spectra at the Fermi energy spread over 2D Brillouin zone. Spin- and $k_\parallel$-resolved transmission probabilities for (a)-(c) Ta-Ta, (d)-(f) W-W and (g)-(i) Mo-Mo MTJs are shown. $\Gamma$, X and M in (a) are the high-symmetry points in ($001$)-plane along which the spectra are calculated. IRS in (b) indicates transmission due to interfacial resonant states corresponding to minority-spins states. Spins are not decomposed for transmission with SOC in (c), (f) and (i). The color bar indicates transmission coefficient plotted on a logarithmic scale.
}
\label{fig:Fig5}
\end{figure}

The transmission at $\Gamma$-point i.e. $k_\parallel=(0,0)$ in the P$^{\uparrow \uparrow}$ channel helps in determining the dominant contributor to the transmission but fails to explain the underlying mechanism of the conductance channels, as in the case of Mo-Mo. As follows from eq.\ref{eq:1}, the conductance is not dominated merely by the transmission at $k_\parallel=(0,0)$ but instead is the integration over the $k_\parallel=k_x,k_y$ plane. To gain further insight into the behavior of conductance, we analyze the transmission coefficient as a function of transverse momentum $k_\parallel(k_x,k_y)$ at the Fermi energy, without and with SOC. The transmission spectra shown in Fig.\ref{fig:Fig5}(a)-(i) are plotted along the (001)-plane in the two-dimensional Brillouin zone (2DBZ) for parallel configuration. The conductance G$_{AP}^{\uparrow \downarrow}$ and G$_{AP}^{\downarrow \uparrow}$ in Ta-Ta, W-W and Mo-Mo overlap each other in Fig.\ref{fig:Fig2}(a). As per the conventional theory of $\emph{Butler et al.}$ \cite{butler2001spin}, the distribution of transmission in $k_\parallel$-space can be easily understood if discussed by means of propagating Bloch wave symmetries like $\Delta_1$, $\Delta_5$, $\Delta_2’$ or $\Delta_2$. In MTJs with Ta, W or Mo capping, explaining transmission with the help of Bloch wave symmetries appears to be feasible as $\emph{Butler’s}$ theory is meant for lattices with square symmetry, unlike IrMn. IrMn, however, has a tetragonal lattice. Therefore, accordingly and as mentioned earlier, for the sake of comparison, we proceed our discussion involving orbital characters rather than wave symmetries. One must note that in Ta-Ta, W-W or Mo-Mo, unlike in IrMn-IrMn, the $\Delta_1$ symmetry is still composed of $\emph{s}$,$\emph{p}{_z}$,$\emph{d}{_{z^2}}$ orbitals and $\Delta_5$ is composed of $\emph{p}{_x}$,$\emph{p}{_y}$,$\emph{d}{_{xz}}$,$\emph{d}{_{yz}}$,  and similarly so on for $\Delta_2’$ and $\Delta_2$.  

In Ta-Ta, transmission patterns of moderate magnitudes are distributed over the whole BZ including the $\Gamma$-point in P$^{\uparrow \uparrow}$ channel as seen in Fig.\ref{fig:Fig5}(a). Whereas in P$^{\downarrow \downarrow}$, in Fig.\ref{fig:Fig5}(b), comparatively high transmission peaks are seen away from the $\Gamma$-point, for example, at $k_{\parallel}$=0.70,0.64 ${\pi/a}$ \cite{zhou2016large}. The cause of these peaks at specific $k_{\parallel}$ is the resonance of interfacial states while that of their large magnitude is the increase in the interfacial states. This increase of interfacial states is due to the large lattice-mismatch-induced strain \cite{cai2012effect}, as explained above.  Therefore, high transmission peaks in P$^{\downarrow \downarrow}$ channel, in contrary to the P$^{\uparrow \uparrow}$ channel validates why  G$_{P}^{\downarrow \downarrow}$ supersedes G$_{P}^{\uparrow \uparrow}$ (see Fig.\ref{fig:Fig2}(a)).  Simultaneously, we inspect the SOC-induced conductance in Fig.\ref{fig:Fig5}(c). The mixing of majority- and minority-spins channel occurs as one cannot decompose spins when SOC is included.  Because of spin-mixing, the Bloch wave symmetries and, therefore, the orbitals tend to mix \cite{lu2012spin}, densely distributing the transmission patterns over the BZ.  Additionally, the magnitude of transmission at normal incidence ($k_{\parallel}=(0,0)$) appears to have increased. When SOC is switched-on, the degeneracy is lifted for the out-of-plane orbital characters, specifically, \emph{d}$_{z^2}$, \emph{d}${_{xz}}$ and \emph{d}${_{yz}}$ \cite{ yang2011first}. The lifting creates additional bands with corresponding orbital characters. This increase in the number of out-of-plane orbital bands increases their contribution which reflects as an increase in the concentration of out-of-plane transmission at $k_{\parallel}=(0,0)$. In W-W, see Fig.\ref{fig:Fig5}(d), a broad high transmission peak appears at $k_{\parallel}=(0,0)$ for P$^{\uparrow \uparrow}$ and also spreads around $k_{\parallel}=(0,0)$. It illustrates the dominant contribution of both, $\Delta_1$ and $\Delta_5$ symmetries , in our case, primarily from the $\emph{d}{_{z^2}}$, $\emph{d}{_{xz}}$ and \emph{d}${_{yz}}$ orbitals as also explained in Appendix D.  Now, since the $\emph{d}{_{xz}}$ and $\emph{d}{_{yz}}$ ($\Delta_5$) are located in the majority-spins channel their contribution in the minority-spins channel is equally reduced. As seen in Fig.\ref{fig:Fig5}(e), a very narrow and single transmission peak (dark red) appears at $k_{\parallel}$=0.67,0.67 ${\pi/a}$, unlike in Ta-Ta, as a result of which G$_{P}^{\downarrow\downarrow}$ in Fig.\ref{fig:Fig2}(a) relatively decreases.  Upon inclusion of SOC, not only that the transmission at $k_{\parallel}$=0,0 intensifies but also the transmission around $k_{\parallel}$=0,0 increases as in Fig.\ref{fig:Fig5}(f). Moreover, the narrow high transmission peak (dark red) located in the minority-spins channel now disappears probably because $\Delta_5$ ($\emph{d}{_{xz,yz}}$) states form an additional conduction channel together with the $\Delta_1$ states at $\Gamma$-point \cite{lu2012spin} .  This formation is brought about by the strong SOC-induced mixing of spins and orbitals. In Fig.\ref{fig:Fig5}(g), for Mo-Mo, a blue dot centered at $\Gamma$-point indicates absence of $\Delta_1$ ($\emph{d}{_{z^2}}$) states whereas a high transmission peak around $\Gamma$ indicates dominating contribution of $\Delta_5$ ($\emph{d}{_{xz,yz}}$) states in P$^{\uparrow \uparrow}$ (see Appendix D).  This relatively high transmission peak which broadens over a considerable area of BZ results into a large integrated transmission coefficient leading to a comparatively increased G$_{P}^{\uparrow \uparrow}$ in Fig.\ref{fig:Fig2}(a).  As the conducting channel in Mo-Mo is strongly governed by $\emph{d}{_{xz,yz}}$ ($\Delta_5$-states), broader high transmission peaks are obvious to occur in the minority-spins channel (see Fig.\ref{fig:Fig5}(h)) which is, anyway,  typically dominated by the $\Delta_5$-states. In consequence, the conductance G$_P^{\downarrow\downarrow}$ and G$_{AP}^{\uparrow \downarrow/\downarrow \uparrow}$ show corresponding increase in Fig.\ref{fig:Fig2}(a). Accordingly, the SOC-induced enchantment of transmission at $\Gamma$-point in Fig.\ref{fig:Fig5}(i) effaces the blue dot in Fig.\ref{fig:Fig5}(g). Unlike in W-W, the broad P$^{\downarrow\downarrow}$ peak associated with the contribution of $\Delta_5$-states in Fig.\ref{fig:Fig5}(h) is left undisturbed in Fig.\ref{fig:Fig5}(i) despite including SOC. This highlights the impact of weak SOC in Mo capping and hence the weak SOC-induced spin-mixing. Furthermore, it is interesting to observe the SOC-induced change in the shape of transmission pattern centered around $k_{\parallel}=(0,0)$ in Fig.\ref{fig:Fig5}(c), (f) and (i) with respect to Fig.\ref{fig:Fig5}((a), (d) and (g). We regard this change to the strength of spin-orbit coupling in the capping layers where SOC in W$>$Ta$>$Mo. Clearly, a moderate change of transmission pattern is observed in Ta-Ta with small notches appearing in the outer circular ring whereas, a remarkably different pattern is observed in W-W MTJ. Mo-Mo shows no change in transmission pattern.

\subsubsection{MTJ with IrMn capping layer}

\begin{figure}[h!]
\centering
\includegraphics[width = 1.0\columnwidth]{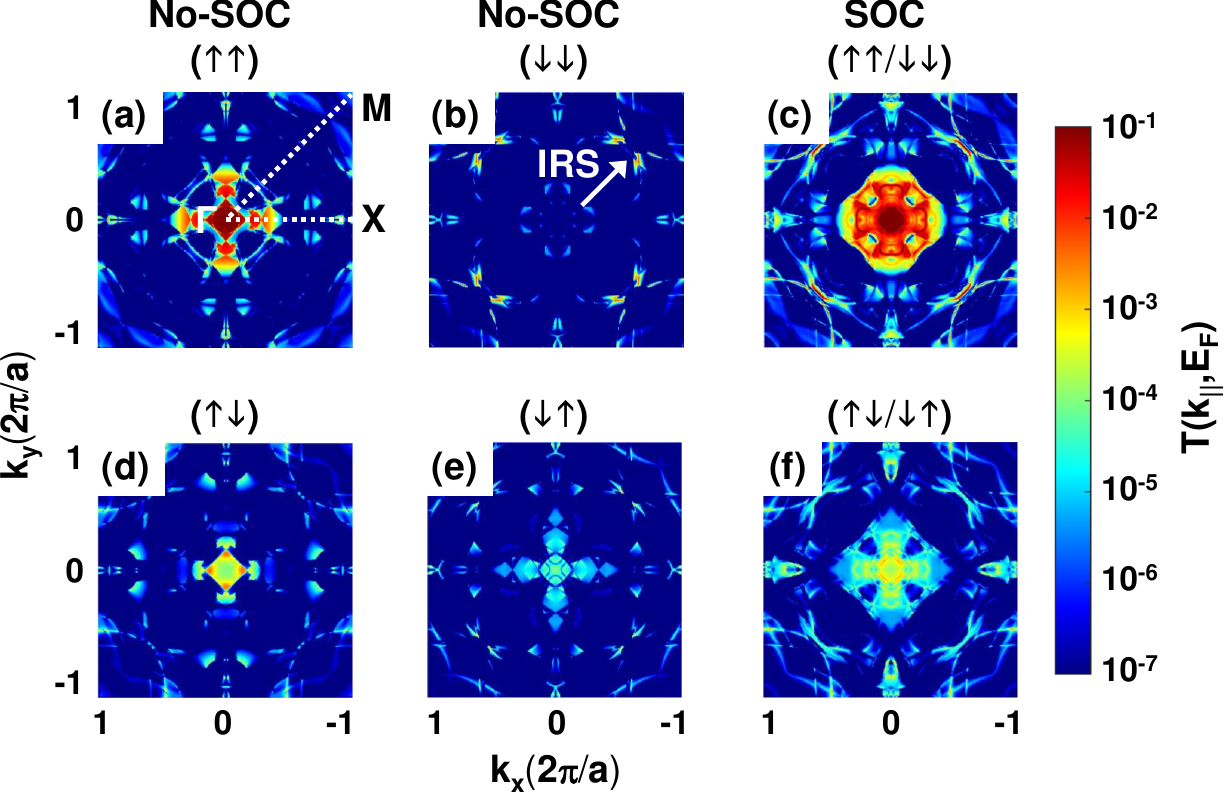}
\caption{Spin- and $k_\parallel$-resolved transmission spectra at the Fermi energy spread over 2D Brillouin zone is shown for IrMn-IrMn MTJ. $\Gamma$, X and M in (a) are the high-symmetry points in ($001$)-plane along which the spectra are calculated. IRS in (b) indicates transmission due to interfacial resonant states corresponding to minority-spins states. Spins are not decomposed for transmission with SOC in (c) and (f). Transmission without SOC and with SOC are shown for (a)-(c) parallel configuration and (d)-(f) anti-parallel configuration. Spins are not decomposed for transmission with SOC in (c) and (f). The color bar indicates transmission coefficient plotted on a logarithmic scale.
}
\label{fig:Fig6}
\end{figure}

Further, we suspect that, unlike Mo, IrMn assists FeCo in maintaining its electronic properties along the ($001$)-direction as in Ref.\cite{kohn2013antiferromagnetic, yang2006exchange}. This, favoured crystallographic orientation, along with an ideal lattice-mismatch, together, must help a propagating wave to transmit unhindered at $\Gamma$. Therefore, without SOC, a relatively concentrated transmission is seen at $k_{\parallel}=(0,0)$ in P$^{\uparrow \uparrow}$ channel in Fig.\ref{fig:Fig6}(a). The transmission coefficient as large as $9.8\times10^{-2}$ is obtained at the $\Gamma$-point  which is governed by the $\emph{d}_{z^2}$ orbital character as detailed in Appendix D. The transmission in P$^{\uparrow \uparrow}$ channel is entirely distributed in the central area of the 2DBZ, unlike in Fig.\ref{fig:Fig5}(g). Therefore, the conductance which is the integration of transmission over \emph{k}-space shows a drop in G$_{P}^{\uparrow \uparrow}$ in Fig.\ref{fig:Fig2}(a). Due to the strong contribution, primarily, from $\emph{d}{_{z^2}}$ orbital character than the $\emph{d}_{xz}$ or $\emph{d}_{yz}$ characters along the (001)-direction, a sparse distribution of transmission patterns all over the BZ is seen. Except that a few sharp but narrow peaks (dark red) are observed at specific $k_\parallel$ such as 0.75,0.63 ${\pi/a}$ in the P$^{\downarrow \downarrow}$ channel in Fig.\ref{fig:Fig6}(b)), unlike in Fig.\ref{fig:Fig5}(h)). When SOC is switched-on, see Fig.\ref{fig:Fig6}(c), firstly the concentration of transmission at and around the $\Gamma$-point is increased with a transmission coefficient of $1.7\times10^{-1}$ at $\Gamma$-point, due to the SOC-induced spin- and orbital-mixing. Second, the transmission pattern in the central area of BZ has completely changed similar to Fig.\ref{fig:Fig5}(e) signifying the impact of strong strength of SOC in the IrMn capping layers. Interestingly, unlike in Fig.\ref{fig:Fig5}(f), the narrow peaks at $k_\parallel$= 0.75,0.63 ${\pi/a}$ do not disappear in Fig.\ref{fig:Fig6}(c) by the SOC effect. This means that the interfacial states-induced \emph{resonant tunneling} which further improves the TMR ratio is preserved in IrMn-IrMn MTJ, unlike in W-W MTJ. Additionally, the transmission patterns distributed in the areas of BZ other than the central region also appear to have intensified. As a result, G$_{P}^{\uparrow \uparrow}$ shows a linear increase over Mo-Mo MTJ in Fig.\ref{fig:Fig2}(c). In the antiparallel configuration, the distinct transmission patterns observed close to the normal incidence in {AP}$^{\uparrow \downarrow}$ and {AP}$^{\downarrow \uparrow}$ channels in Fig.\ref{fig:Fig6}(d) and Fig.\ref{fig:Fig6}(e), respectively, justify the non-identical antiparallel conductance in Fig.\ref{fig:Fig2}(a).  As a consequence, the total G$_{AP}$ (${\uparrow \downarrow} + {\downarrow \uparrow}$) remains low producing a large TMR. Here, we strongly anticipate that the antiferromagnetic ordering of IrMn administers the transport. This is because we allow only the spins of one of the FeCo electrode to rotate to achieve antiparallel alignment but prevent IrMn from changing. Due to the uncompensated spin arrangement between AFM-IrMn and FM-FeCo a magnetically imbalanced interface is created which breaks the symmetry.  Switching on SOC refrains from spin-selectivity and hence no non-identical signatures can be found. As well, a broad transmission pattern is seen around $k_{\parallel}=(0,0)$ since the magnetically imbalanced interfaces no longer exist. 

\subsection{Effect of SOC on Resonant Tunneling}
\begin{figure}[h!]
\centering
\includegraphics[width = 0.7\columnwidth]{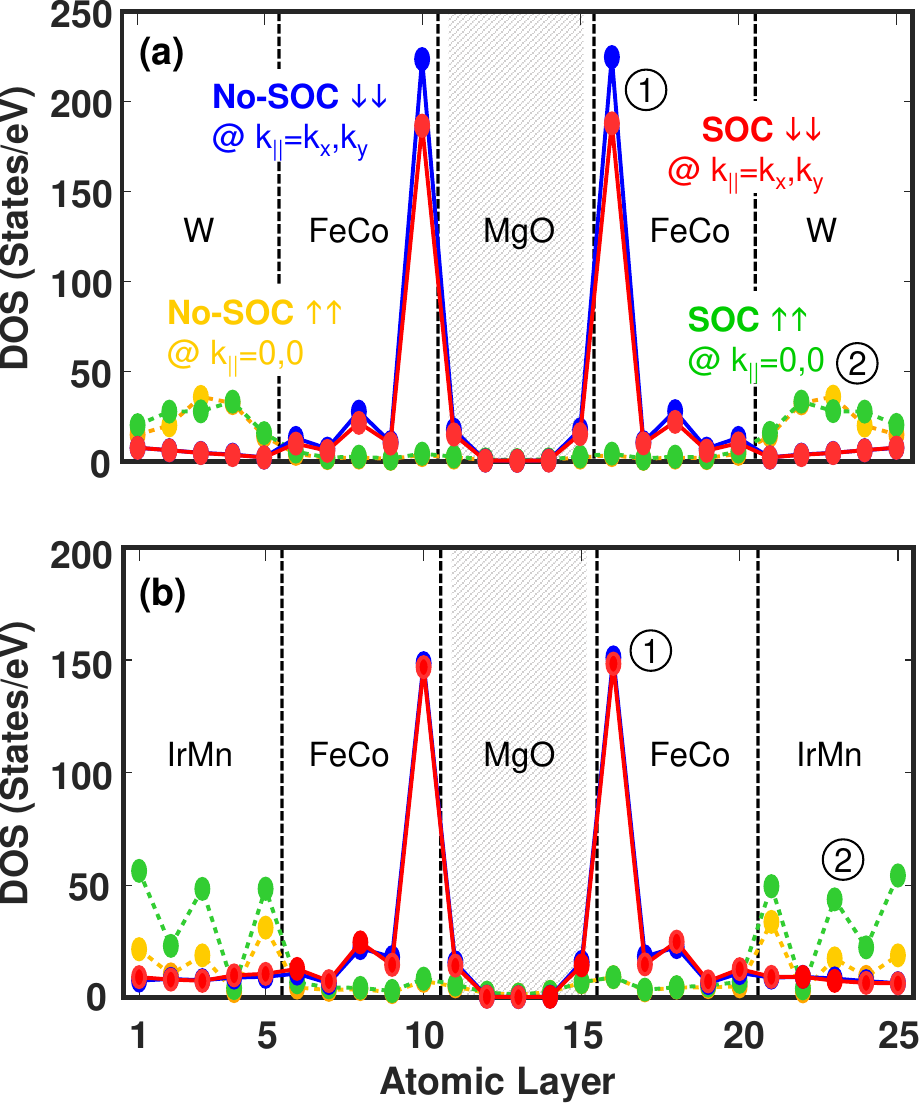}
\caption{Layer-resolved density-of-states calculated for P$^{\uparrow \uparrow}$ channel at $k_{\parallel}=(0,0)$ (yellow, green) and for P$^{\downarrow \downarrow}$ channel at $k_{\parallel}=(k_x,k_y)$ (red, blue) in (a) W-W and (b) IrMn-MTJ, without and with SOC. The numbers inside the circle highlight the change occurring upon switching on SOC. Higher density peaks (red, blue) at the MgO interface contribute to the transport called 'Resonant tunneling'.}
\label{fig:Fig7}
\end{figure}

Understanding the contribution of interfacial resonant states (IRS) to the transport  is crucial as it further improves the TMR ratio \cite{butler2001spin,lu2012spin,zhou2016large}. Without SOC, IRS are located in P$^{\downarrow \downarrow}$ channel which is dominated by $\emph{d}{_{xz,yz}}$ ($\Delta_5$) orbitals. Understanding the role of IRS upon switching on SOC, however, becomes complex because of the spin-mixed conducting channels. When SOC is switched on, on one hand, the transmission at $k_{\parallel}=(k_x,k_y)$ in P$^{\downarrow \downarrow}$ channel, where the IRS are located, is reduced because of the SOC-induced Rashba effect\cite{popescu2005influence,chantis2007tunneling,lu2012spin}. On the other hand, simultaneously, the transmission at normal incidence at $k_{\parallel}=(0,0)$ in P$^{\uparrow\uparrow}$ channel slightly improves, probably, because of an increased contribution from the additional $\emph{d}{_{z^2}}$ ($\Delta_1$) bands built by SOC-induced lifting of spin degeneracy. Typically, the concentration of transmission at $k_{\parallel}=(0,0)$ in P$^{\uparrow\uparrow}$ channel and at $k_{\parallel}=(k_x,k_y)$ in P$^{\downarrow \downarrow}$ channel dominate the conductance in parallel configuration and, eventually, the TMR. Therefore, we examine the spin-decomposed density-of-states (DOS), without and with SOC, both at $k_{\parallel}=(k_x,k_y)$ in P$^{\downarrow \downarrow}$ and $k_{\parallel}=(0,0)$ in P$^{\uparrow\uparrow}$ as plotted in Fig.\ref{fig:Fig7} to investigate above mentioned arguments. In Fig.\ref{fig:Fig7}(a) and (b) we illustrate DOS for W-W and IrMn-IrMn MTJs, respectively. Besides, as deduced from Fig.\ref{fig:Fig5} and \ref{fig:Fig6}, W-W and IrMn-IrMn are best suitable MTJs to discuss the effect of SOC. The states at the FeCo/MgO interface in Fig.\ref{fig:Fig7}(a) at $k_{\parallel}=(k_x,k_y)$ seem to have reduced upon switching on SOC whereas no significant change occurs at $k_{\parallel}=(0,0)$. In contrast, no significant change occurs at $k_{\parallel}=(k_x,k_y)$ Fig.\ref{fig:Fig7}(b) while a significant change is seen in IrMn capping layers at $k_{\parallel}=(0,0)$. From these facts we can infer that: (a) in IrMn-IrMn MTJ the IRS-induced resonant tunneling is preserved in P$^{\downarrow \downarrow}$ channel, despite strong SOC and, in parallel, (b) the contribution of IrMn capping layer to the transmission in P$^{\uparrow\uparrow}$ channel is developed. For these reasons, we speculate that IrMn-IrMn MTJ proves slightly more efficient system as compared to W-W and, inevitably, than other MTJs. 

\subsection{Scattering in IrMn-IrMn MTJ}
\begin{figure}[h!]
\centering
\includegraphics[width = 1.0\columnwidth]{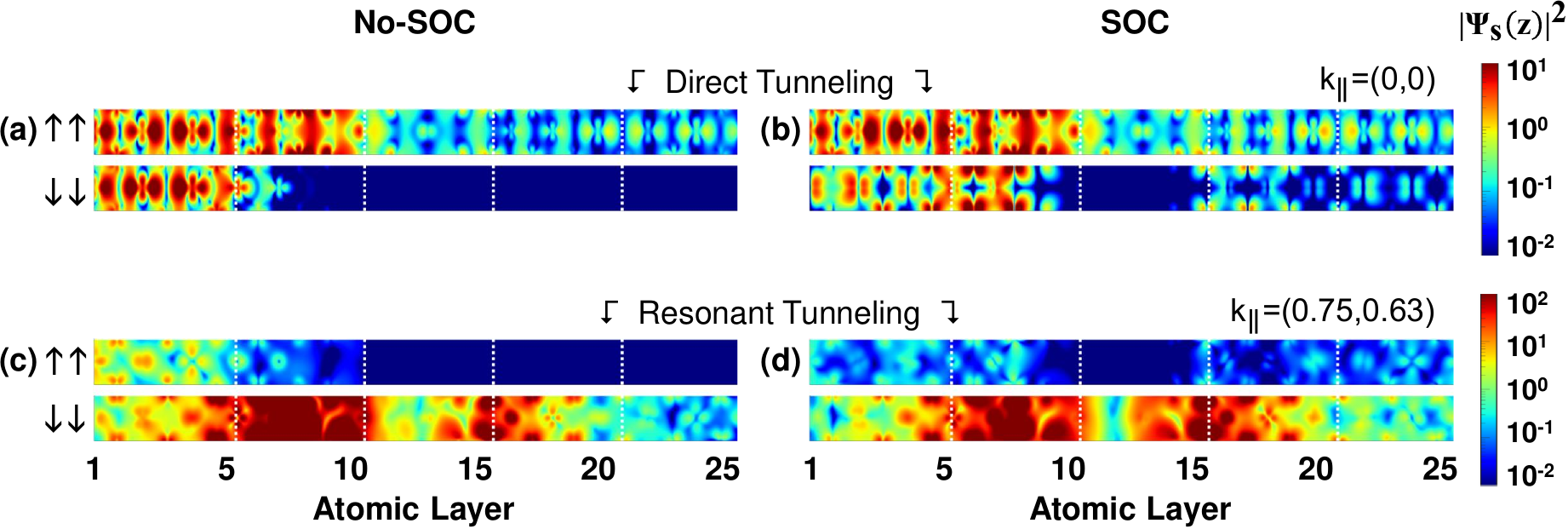}
\caption{Real-space distribution of squared modulus of scattering states ${|\Psi_s(z)|^2}$ in parallel configuration across the IrMn-IrMn MTJ, under equilibrium, at the Fermi energy. The scattering states are calculated at different $k_{\parallel}$  points: (a),(b) $k_{\parallel}=(0,0)$ and (c),(d) $k_{\parallel}=(0.75,0.63)$, without and with SOC. The values of $k_{\parallel}$  are given in the units of $2\pi/a$. In (a) P$^{\uparrow\uparrow}$ indicates 'direct tunneling' mechanism at $\Gamma$-point, while in (b) P$^{\downarrow\downarrow}$ indicates \emph{Resonant tunneling} governed by IRS. The vertical dashed lines roughly mark the interfaces. The color bar represents logarithmic scaled $|\Psi_s(z)|$  when an electron travels along the direction of transport (z-direction).}
\label{fig:Fig8}
\end{figure}

Emergence of additional conducting channels upon inclusion of SOC can be easily visualized by means of real-space distribution of scattering states, ${|\Psi_s(z)|}$. In addition, with analysis of scattering states one can understand the path taken by a propagating electron wave in different conducting channels. For better insight, we show the spin-decomposed distribution, before and after including SOC, at $k_{\parallel}$=0,0 in Fig.\ref{fig:Fig8}(a) and (b) and $k_{\parallel}=(k_x,k_y)$ in Fig.\ref{fig:Fig8}(c) and (d), in IrMn-IrMn MTJ. The scattering states distribution is obtained from a squared modulus of the right incident wave, ${|\Psi_s(z)|^2}$, traveling from the left capping layer towards the right capping. At $\Gamma$-point ($k_\parallel=0,0$), the right incident wave traveling from the left IrMn capping layer enters the left FeCo-electrode and tunnels through the MgO-barrier with a delayed exponential decay indicating a direct tunneling mechanism in P$^{\uparrow \uparrow/\downarrow\downarrow}$ channels (see Fig.(\ref{fig:Fig8}(a)). In P$^{\downarrow\downarrow}$, the wave decays rapidly and cannot propagate to the other side of the barrier.  The scattering wave in  P$^{\uparrow\uparrow}$ in Fig.(\ref{fig:Fig8}(a) and (b) is majorly composed of \emph{d}$_{z^2}$ orbital character of Ir, Mn (Fe, Co) and \emph{p}$_z$ character of O in the MgO-barrier.  On inclusion of SOC, in Fig.(\ref{fig:Fig8}(b), the contribution of \emph{d}$_{z^2}$ in P$^{\uparrow\uparrow}$ appears relatively stronger which is attributed to the SOC-induced increase of \emph{d}$_{z^2}$ band levels due to lifting of degeneracy. At the same time, an obvious change in the dominating orbital character of the scattering wave in P$^{\downarrow\downarrow}$ channels is seen in Fig.(\ref{fig:Fig8}(a) and (b). Contrastingly, when SOC is switched on,  P$^{\downarrow\downarrow}$ channel is dominated by the $\emph{d}{_{xz,yz}}$ orbital characters and emerge on the other side of the MgO-barrier. are seen to . Here, one can see an additional conducting channel of $\emph{d}{_{xz,yz}}$ is formed which tunnels towards the right IrMn layer. This formation, as mentioned earlier, is attributed to the SOC-induced orbital-mixing because of which the $\emph{d}{_{xz,yz}}$ character starts to transmit at normal incidence i.e. at $\Gamma$-point. Important to note, as $\emph{d}{_{xz,yz}}$ characters are compatible with spin-down configuration, their contribution is precisely observed in the P$^{\downarrow\downarrow}$ channel,only. The darker blue region within the MgO-barrier in P$^{\downarrow\downarrow}$ in Fig.(\ref{fig:Fig8}(b) manifests rapid decay of $\emph{d}{_{xz,yz}}$. This further justifies the influence of SOC-induced co-existence of \emph{d}$_{z^2}$ and $\emph{d}{_{xz,yz}}$ orbital contribution in transmission spectrum in Fig.\ref{fig:Fig6}(c) resulting into tremendously large TMR.

Next, we proceed with the explanation of scattering wave at $k_{\parallel}=(k_x,k_y)$, in Fig.\ref{fig:Fig8}(c) and (d), without and with SOC, respectively. Here, we first discuss P$^{\downarrow\downarrow}$ channel in which IRS appear at specific $k_{\parallel}=(k_x,k_y)$ . As one can see, in P$^{\downarrow\downarrow}$, the scattering wave is significantly dominated by the interfacial resonant states (IRS) localized at FeCo/MgO interface. This scattering wave propagates with almost negligible attenuation and appears on the other side of the barrier. As a result, the contribution of scattering states is found to be extremely large around the MgO-barrier interface. These states travel by coupling with the evanescent states within the MgO-barrier indicating a resonant tunneling mechanism in P$^{\downarrow\downarrow}$ channel. More importantly, switching on SOC does not hinder the contribution of IRS which continues to dominate the P$^{\downarrow\downarrow}$ channel in IrMn-IrMn MTJ as shown in Fig.\ref{fig:Fig8}(d), unlike in W-W MTJ.  Besides, a new conducting channel is found to establish in P$^{\uparrow\uparrow}$ of Fig.\ref{fig:Fig8}(d), unlike W-W MTJ. A contrasting results of the scattering wave in W-W MTJ which presents better understanding of Fig.(\ref{fig:Fig7}(a) are depicted in Appendix F.

\subsection{Bias-Dependent TMR in IrMn-IrMn MTJ}
Finally, in Fig.\ref{fig:Fig10}, we discuss non-equilibrium transport in IrMn-IrMn MTJ. The calculated total current as a function of bias voltage, without and with SOC, is presented for P (I$_P^{tot}$) and AP (I$_{AP}^{tot}$) configurations in Fig.\ref{fig:Fig10} (a) and (b), respectively. The total current, without SOC, is the sum of spin-decomposed currents, for example, I$_P^{tot}$=I$^{\uparrow\uparrow}$+I$^{\downarrow\downarrow}$ for P configuration. Corresponding spin-decomposed currents, without SOC, are shown within the insets. In parallel configuration, the No-SOC I$_P^{tot}$ which is governed by I$^{\uparrow \uparrow}$ increases linearly in the bias range of 0 to 0.05 V. I$^{\downarrow\downarrow}$, however, remains majorly low. Similarly, the No-SOC I$_{AP}^{tot}$ shows a linear increase which is mostly contributed by I$^{\uparrow\downarrow}$. Unconventionally, the distinct response of I$^{\uparrow\downarrow}$ and I$^{\downarrow\uparrow}$ seen in the inset of Fig.\ref{fig:Fig10}(b) arises from the non-identical behavior of antiparallel channels in IrMn-IrMn, explained earlier.  As a result, I$_{AP}^{tot}$ of nearly two orders of magnitude smaller than I$_P^{tot}$ is obtained. This huge difference in I$_P^{tot}$ and I$_{AP}^{tot}$ leads to a high TMR value. In Fig.\ref{fig:Fig10}(c), the non-equilibrium TMR as a function of current is defined as TMR=(I$_P^{tot}$-I$_{AP}^{tot}$)/I$_{AP}^{tot}$. In our calculations, the TMR at zero bias is determined by means of transmission coefficient due to vanishing currents at 0 V \cite{waldron2007ab}. The TMR, without SOC, descends gradually with increasing bias voltage. When SOC is switched-on, the magnitude of I$_P^{tot}$ for all finite voltages appears to have increased with, beneficially, no change in I$_{AP}^{tot}$.  Accordingly,  an increase in TMR, with SOC, is observed with a rapid decay until 0.03 V. 

\begin{figure}[h!]
\centering
\includegraphics[width = 0.5\columnwidth]{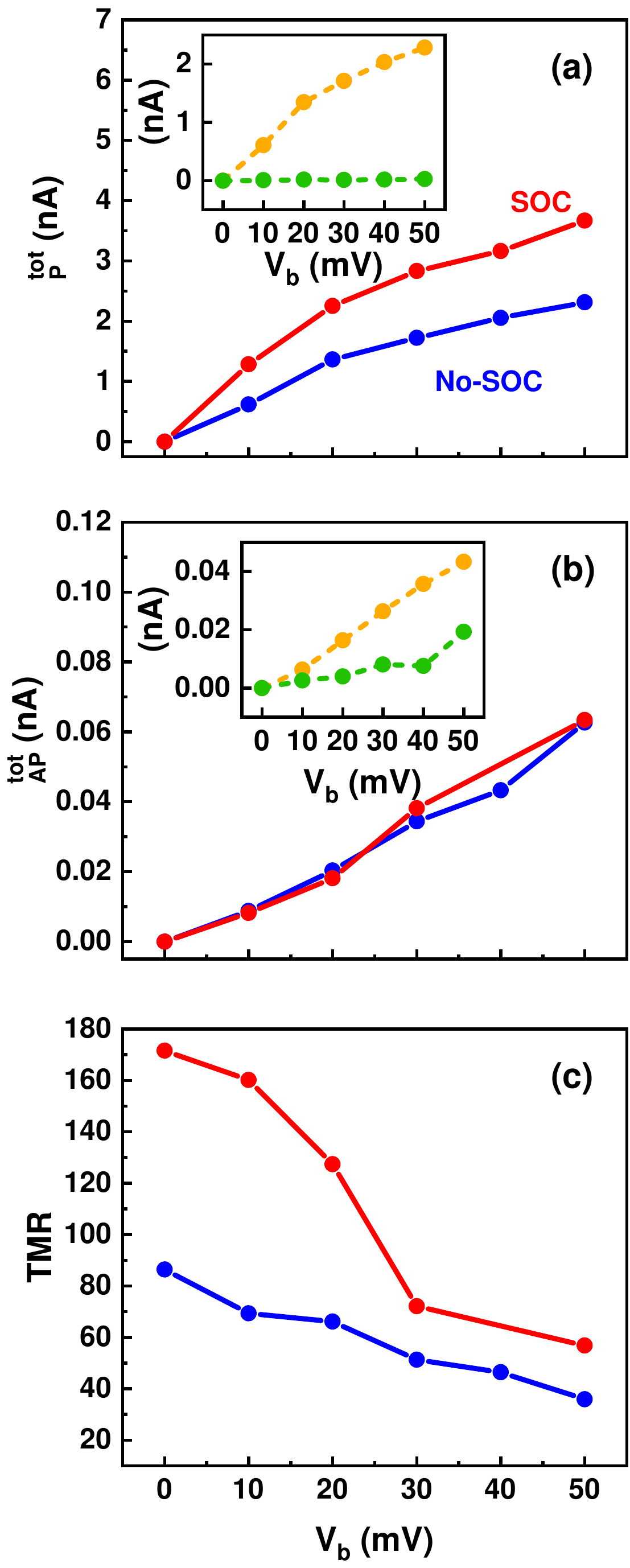}
\caption{Bias-dependent transport properties for IrMn-IrMn MTJ. (a), (b) Total current, without and with SOC, in P and AP spin configuration, respectively. Inset of (a), (b): spin-decomposed current (yellow, green), without SOC. (c) TMR without (blue) and with SOC (red).}
\label{fig:Fig9}
\end{figure}

\section{Conclusion}
To summarize, we probe into two aspects which are considered substantial pertaining to the tunneling in a magnetic junction- (i) the spin-orbit coupling strength and (ii) the magnetic property of the capping layers. Theoretically, a comparative study of transport, without and with SOC, is presented  by employing different capping layers such as non-magnetic Ta, W and Mo along with an antiferromagnetic L1$_0$-IrMn. It is shown that, without SOC, a largest magnitude of TMR is obtained in IrMn-IrMn MTJ (up to 8640$\%$), despite, it’s parallel conductance (G$_{P}^{\uparrow \uparrow/\downarrow \downarrow}$) being smaller in comparison with the Mo-Mo MTJ. This peculiarity of TMR is attributed to: (a) a relatively concentrated transmission at the normal incidence ($\Gamma$-point) in P$^{\uparrow \uparrow}$ and (b) the non-identical behavior of {AP}$^{\uparrow \downarrow}$ and {AP}$^{\downarrow \uparrow}$ resulting into a lowest antiparallel conductance (G$_{AP}^{\uparrow \downarrow}$ + G$_{AP}^{\downarrow \uparrow}$), in IrMn-IrMn MTJ. This distinct signatures of AP channels is shown to have originated from the antiferromagnetic property of IrMn capping layer which creates a magnetically imbalanced AFM/FM interface.

Upon switching on SOC,  firstly, the TMR ratio is found to have increased in all MTJs with a particularly pronounced increase in W-W and IrMn-IrMn MTJs. This signifies the importance of utilizing capping layers with strong SOC. In addition, in IrMn-IrMn MTJ,  the difference between the TMR ratio, without and with SOC, is $\sim$25\% larger in comparison to W-W. Apparently, this giant difference in IrMn-IrMn is brought about by: (a) the enhancement in parallel conductance (G$_P$) over all other MTJs, unlike conductance without SOC (G$_{P}^{\uparrow \uparrow/\downarrow \downarrow}$) and (b) the antiparallel conductance (G$_AP$) which continues to remain low even after SOC is switched-on.  This enhancement of G$_P$ is further investigated by calculating the scattering wave function ${|\Psi_s(z)|}$ propagating within the IrMn-IrMn MTJ, without and with SOC. It is seen that, SOC-induced lifting of degeneracy increases transmission in P$^{\uparrow \uparrow}$ whereas the spin- and orbital-mixing increases transmission in  P$^{\downarrow \downarrow}$, at the $\Gamma$-point ($k_{\parallel}$=0,0). Interestingly and unconventionally, it appears that the interfacial resonant states at specific $k_{\parallel}=(k_x,k_y)$ also intensively contribute to the transmission, when SOC is included, preserving the contribution of resonant tunneling to G$_P$. Furthermore, to support these arguments regarding transport in IrMn-IrMn MTJ, we perform non-equilibrium bias-dependent calculations in the range of 0-0.05V for IrMn-IrMn MTJ. These IV-results demonstrate that, when SOC is included, the tunneling current in parallel configuration (I$_P^{tot}$) is enhanced for every bias voltage whereas, beneficially, the antiparallel current (I$_{AP}^{tot}$) remains almost unchanged. As a result, the SOC-induced TMR is always larger than the TMR without SOC, for every bias voltage. However, this TMR descends with increasing bias voltage.  

\section*{Acknowledgement}

We would like to thank Dr. Yu Feng and Jiaqi Zhou for their valuable discussions on modeling with Nanodcal package. The authors gratefully acknowledge the National Key R$\&$D Program of China (No. 2018YFB0407602), the Beijing Municipal Science and Technology Project (No. Z201100004220002), the International Collaboration Project (No. B16001), and the Beihang Hefei Innovation Research Institute (Project No. BHKX-19-02) for their financial support of this work. Dr. B Dup\'e acknowledges funding by the DFG under Grant No. DU 1489/3-1. S. Chandrashekhar Koli and Dr. B. Dup\'e gratefully acknowledge the Graduate School of Excellence Materials Science in Mainz (MAINZ) for financial support. Parts of this research were conducted using the supercomputer Mogon and/or advisory services offered by Johannes Gutenberg University Mainz (hpc.uni-mainz.de), which is a member of the AHRP (Alliance for High Performance Computing in Rhineland Palatinate,  www.ahrp.info) and the Gauss Alliance e.V. The authors gratefully acknowledge the computing time granted on the supercomputer Mogon at Johannes Gutenberg University Mainz (hpc.uni-mainz.de) and computing time at Mogon supercomputers.

\appendix

\section{L1${_0}$-IrMn Crystal Structure}
An ordered CuAu-I-type face-centered tetragonal (FCT) L1$_0$-IrMn \cite{pal1968magnetic,fuke2000exchange} is considered as one of the key antiferromagnetic metal in magnetic devices, especially, spin-orbit coupled devices. Here, we employ L1${_0}$-IrMn from $D{_{4h}}$ family and \emph{P4/mmm} (No. 123) space group with lattice constants  $\bold{i}$=$\bold{j}$=3.855 {\AA} and $\bold{k}$=3.643 {\AA} \cite{umetsu2004pseudogap}–\cite{zhang2014spin}. The corresponding first Brilluoin Zone is as shown in Fig.\ref{fig:Fig11}(a).  To obtain better insight into the antiferromagnetic electronic structure of tetragonal IrMn we calculate the band structure \cite{umetsu2004pseudogap} in Fig.\ref{fig:Fig11}(b) with aforementioned orbital character contributions. Mixing of these orbital characters is seen upon inclusion of SOC in Fig.\ref{fig:Fig11} (c). The occupancy of $\emph{d}{_{z^2}}$ character which is the dominant \emph{d}-orbital contributor to the transport is increased in the (001)-plane along the $\Gamma$-X-M-$\Gamma$ lines. Finally, it appears that the $\emph{d}{_{z^2}}$ and $\emph{d}{_{xz}}$ character, both, exclusively contribute to the lifting of degeneracy resulting into increased number of bands with respective occupancies.    

\begin{figure}[h!]
\centering
\includegraphics[width = 1.0\columnwidth]{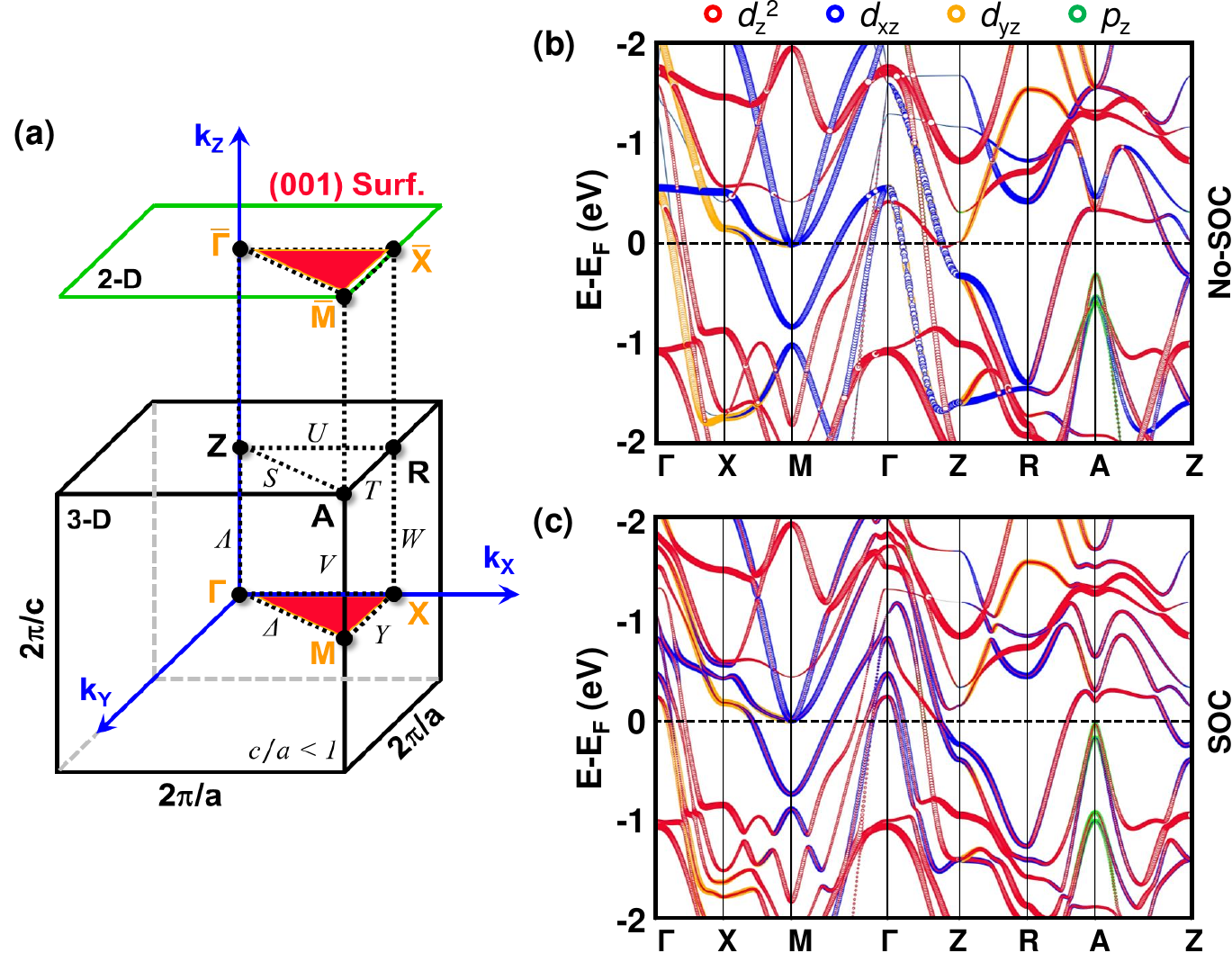}
\caption{(a) Bulk Brillouin zone and the (001) surface projection of a simple tetragonal lattice with symmetry points and lines. (b) and (c), DFT calculated bulk band structure of IrMn without and with SOC, respectively, in the AFM configuration. Bands are projected on out-of-plane  \emph{p}- and \emph{d}-orbitals with the colored spheres indicating the varying contribution of these out-of-plane orbitals. Upon inclusion of SOC, in (c), band-splitting and orbital-mixing can be seen. The Fermi level is located at Energy = 0 eV.}
\label{fig:Fig10}
\end{figure}

\section{Interface Stability}

The chemical stability of an Ir-terminated 5ML-IrMn capping was inspected with adjacent FM electrode in structurally relaxed IrMn($001$)[$110$]/FM($001$)[$100$] bilayer and IrMn($001$)[$110$]/FM($001$)[$100$]/MgO($001$)[$110$] trilayer. Besides, the stability of a FM electrode termination was also tested with respect to both, the IrMn capping and the MgO-barrier layer. In all our calculations, we consider B${_2}$ Iron-Cobalt along (001) direction \cite{zhang2004large,bonell2012spin} sandwiched between a 5ML-X capping and a 5ML-MgO barrier. The total free energy, $E_{free}$ (eV) and the binding energy, E$_b$ (eV.{\AA}$^{-1}$) is computed for Fe- and Co-terminated Iron-Cobalt interfaces. Forthcoming, in this paper,  FeCo refers to the Fe-terminated Iron-Cobalt electrode while CoFe refers to the Co-terminated one. The in-plane lattice of all stacks is fixed to the lattice constant of Iron-Cobalt i.e. 2.83 {\AA}.

The binding energy E$_b$ is defined as \cite{burton2006atomic}, 

\begin{eqnarray}
E_b = E_{ \mathrm{IrMn} }+E_{ \mathrm{FM} }+E_{ \mathrm{MgO} }-E_{ \mathrm{IrMn/FM/MgO} }
\label{eq:4}
\end{eqnarray}

Here, E$_{IrMn/FM/MgO}$ is the total energy of IrMn/FM/MgO  trilayer. E$_{IrMn}$, E$_{FM}$, and E$_{MgO}$ correspond to the total energies of clean IrMn, FM and MgO, respectively, calculated without further relaxation. In Table \ref{table:I}, a large E$_{free}$ obtained for FeCo/MgO bilayer implies that the Fe-atoms prefer to accumulate at the MgO interface than in the bulk which is consistent with the experiments \cite{hindmarch2010fe}. Whereas, the large E$_b$ in CoFe/MgO than in the FeCo/MgO bilayer implies that the Co-terminated interface is energetically more stable than that of the Fe-terminated one. This can be attributed to the strong Co-O bonding around the CoFe/MgO interface \cite{burton2006atomic}. This argument is additionally supported by analyzing the interface bond-length in FeCo(CoFe)/MgO bilayers where Fe-O length is found to be 2.23 {\AA} \cite{burton2006atomic,butler2001spin} while the Co-O length is 2.15 {\AA} \cite{burton2006atomic}, respectively. Therefore, Co-termination in CoFe/MgO-based magnetic junctions has been evidently preferred.

At the same time, IrMn integrated bilayers and trilayers showed a relatively large E$_{free}$ along with a large E$_b$ in Fe-terminated structures. This suggests that: (i) similar to FeCo/MgO, Fe-atoms from the FeCo electrode prefer to reside at the interface than in the bulk and (ii) Ir-Fe termination is energetically more stable than Ir-Co. The Ir-Co interface distance is found to be 1.73 {\AA} while the distance of Ir-Fe is 1.67 {\AA}. This can be because: (a) a strong spin-orbit coupling in Ir results into a strong Ir-Fe interfacial coupling \cite{dupe2016engineering} and (b) IrMn-Fe favors a stable growth in the (001)-direction while IrMn-Co favors (111)-directed growth \cite{kohn2013antiferromagnetic,yanes2013exchange,castro2013role}. Besides, the Fe-Ir bonding helps to increase the interfacial PMA and TMR \cite{nozaki2020voltage}. As we want to emphasize the impact of an AFM-IrMn capping, we employ Fe-terminated FeCo electrode even in structures integrating Ta, W or Mo capping, for the sake of comparison. The favoured in-plane crystallographic arrangement of B${_2}$ IrMn and FeCo interface is illustrated in Fig.\ref{fig:structure}(c) and the longitudinal interfacial stacking along the direction of transport is shown in Fig.\ref{fig:structure}(d). 

\begin{table}[h!]
\caption{The total free energy of fully-relaxed structures and the Binding energy as defined by Eq.\ref{eq:4}. FeCo and CoFe refer to Fe-terminated and Co-terminated interface, respectively.}
\centering
\begin{tabular}{m{3cm} m{2.5cm} m{2.5cm}} 
 \hline\hline \\ 
 Structure & $E_{free}$ (eV) & $E_b$(eV\AA$^{-1}$) \\ [1.5ex]
 \hline \\ 
 FeCo/MgO & -95.243 & 0.568\cite{burton2006atomic}  \\ [0.5ex]
 CoFe/MgO & -94.265 & 0.69\cite{burton2006atomic} \\ [0.5ex]
 IrMn/FeCo & -81.506 & 2.550  \\ [0.5ex]
 IrMn/CoFe & -80.247 & 2.311  \\ [0.5ex]
 IrMn/FeCo/MgO & -140.314 & 3.105  \\ [0.5ex]
 IrMn/CoFe/MgO & -139.195 & 3.008 \\ [1ex] 
 \hline\hline
 \label{table:I}
\end{tabular}
\end{table}

\section{Calculated LDOS of AP$^{\uparrow \downarrow/\downarrow \uparrow}$ channels}
\begin{figure}[h!]
\centering
\includegraphics[width = 0.6\columnwidth]{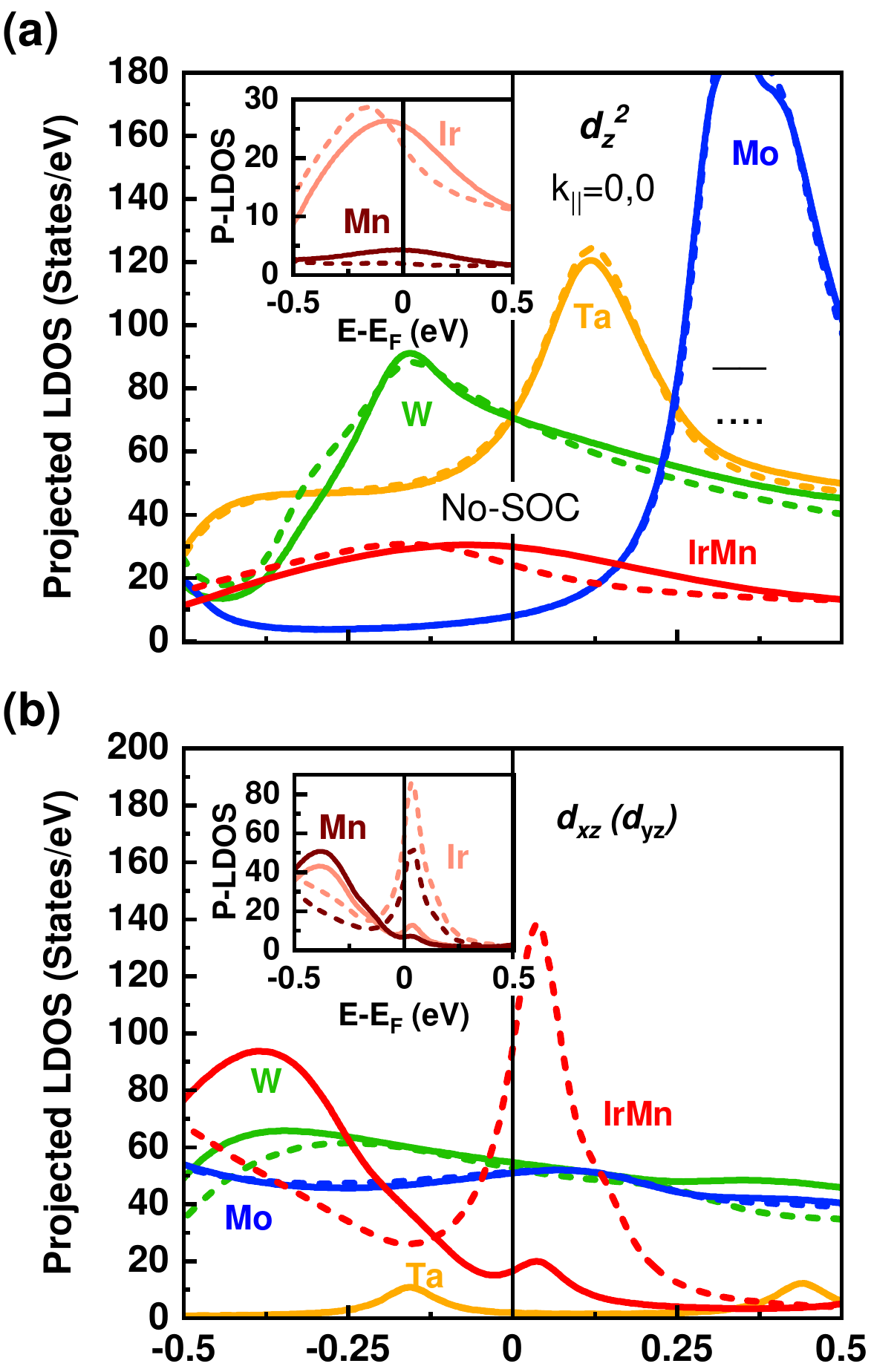}
\caption{In (a) the equilibrium transmission coefficient of all MTJs as a function of electron energy in the antiparallel channels, $\uparrow \downarrow$ and $\downarrow \uparrow$, together with the local DOS projected on (b) $\emph{d}{_{z^2}}$ and (c) $\emph{d}{_{xz,yz}}$ orbital characters of different atomic constituents in capping layers. The calculations are performed without SOC and specifically at $\Gamma$-point.}
\label{fig:Fig11}
\end{figure}

Since the interplay between \emph{d}-orbitals of X and FeCo is of primary importance, we carefully project LDOS onto the \emph{d}-orbitals of X, particularly, only the out-of-plane characters (see Fig.\ref{fig:Fig11}(a) and (b)). In addition to the capping layer as a whole, our concern lies in the contribution of the independent atomic constituents, as well. So, we disintegrate the LDOS by projecting it onto individual atomic species of a capping layer, specifically, in IrMn. In Ta-Ta, W-W and Mo-Mo MTJ, the AP$^{\uparrow \downarrow}$ and AP$^{\downarrow \uparrow}$ channels in Fig.\ref{fig:Fig3} present purely an identical crossing at the Fermi energy which is supported by the identical population of $\emph{d}{_{z^2}}$ or the $\emph{d}{_{xz,yz}}$ orbitals, at the Fermi energy.  In Ta-Ta, majorly $\emph{d}{_{z^2}}$ orbitals contribute to the antiparallel conductance while that in W-W both the $\emph{d}{_{z^2}}$ and $\emph{d}{_{xz,yz}}$ orbitals contribute. The $\emph{d}{_{xz,yz}}$ orbitals which compose $\Delta_5$-states are strongly populated in Mo-Mo MTJ. Conversely in IrMn-IrMn,  AP$^{\uparrow \downarrow}$ and AP$^{\downarrow \uparrow}$ channels manifest completely distinct magnitudes of transmission at the Fermi level (inset of Fig.\ref{fig:Fig3}(a)).  Firstly, both, $\emph{d}{_{z^2}}$ and $\emph{d}{_{xz,yz}}$ orbital characters are considerably populated at the Fermi level. Second, clearly, not only that the atomic species, Ir and Mn, show a non-identical population of states at Fermi level but also their contribution in respective AP$^{\uparrow \downarrow}$ and AP$^{\downarrow \uparrow}$ channels is strongly non-identical. The distinct population of states in Ir and Mn is obvious as they are completely different atomic species with contrasting material properties. However, the distinct magnitudes in AP$^{\uparrow \downarrow}$ and AP$^{\downarrow \uparrow}$ channel points out a performance purely influenced by the magnetic modifications which, in turn, forced us to inspect the behavior of adjacent FeCo (free-layer), as well. After careful examination, we find that even in FeCo free-layer, AP$^{\uparrow \downarrow}$ and AP$^{\downarrow \uparrow}$ channels show a non-identical crossing at Fermi level (LDOS not shown). We anticipate this behavior to be an outcome of a magnetic imbalance between an antiferromagnetic IrMn and a ferromagnetic FeCo either at the interface due to asymmetric interfacial spin-alignment or in the bulk due to induced magnetism. Again, in the antiparallel configuration we adjust only the spin-alignment of FeCo free-layer but leave IrMn unaltered. Also as one knows, Mn is (antiferro) magnetic in itself whereas Ir possesses hybridization-induced magnetization because of the adjacent FeCo \cite{doi2009magnetization}.

\section{Calculated LDOS of P$^{\uparrow\uparrow}$ channel}
\begin{figure}[h!]
\centering
\includegraphics[width = 0.6\columnwidth]{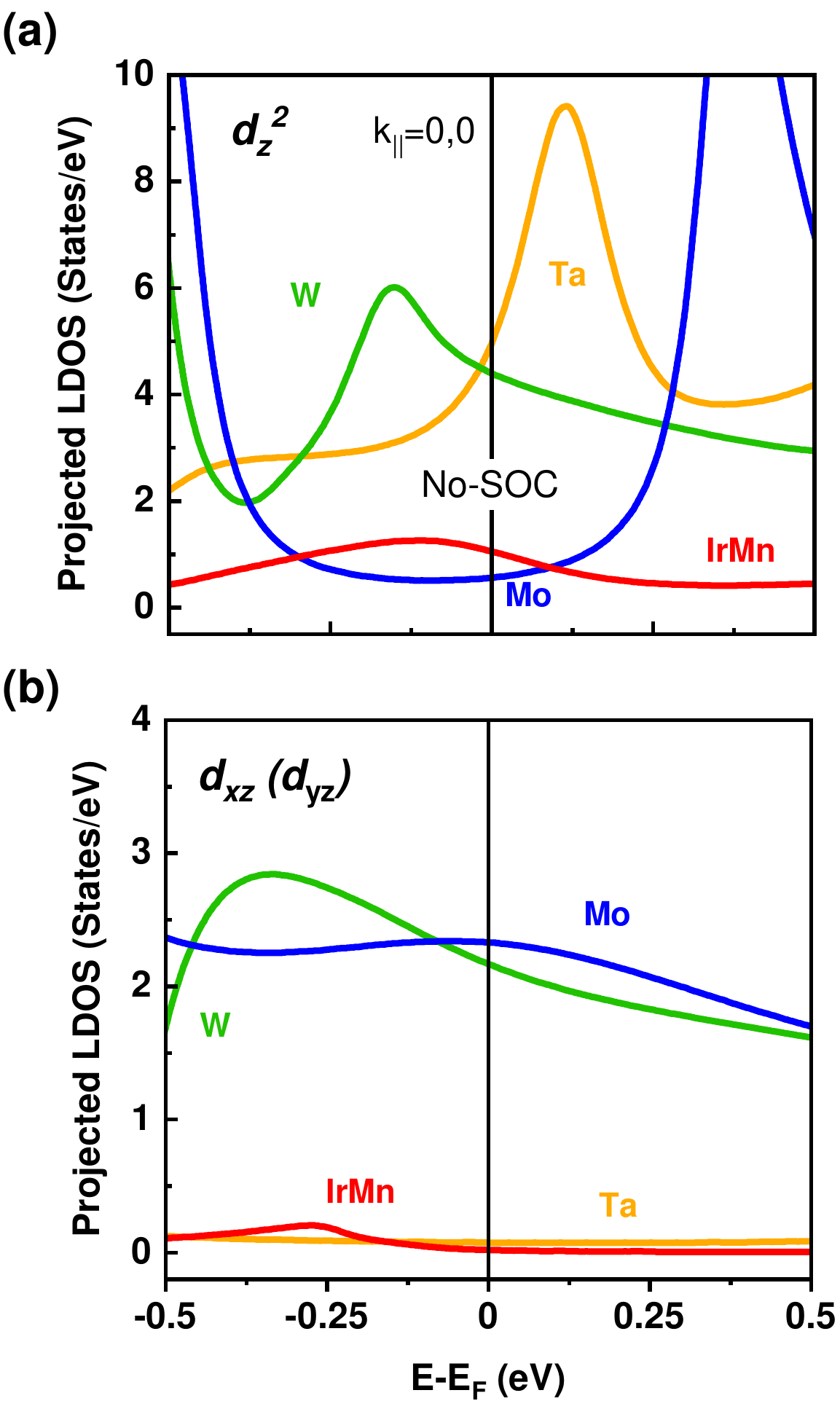}
\caption{In (a) the equilibrium transmission coefficient of all MTJs as a function of electron energy in the majority-spin channel in parallel configuration, $\uparrow \uparrow$ together with the local DOS projected on (b) $\emph{d}{_{z^2}}$ and (c) $\emph{d}{_{xz,yz}}$ orbitals in capping layers. The calculations are performed without SOC and specifically at $\Gamma$-point.}
\label{fig:Fig12}
\end{figure}

Corresponding LDOS of the transmission coefficient presented in Fig.\ref{fig:Fig4} is projected on the $\emph{d}{_{z^2}}$ and $\emph{d}{_{xz,yz}}$ orbitals as plotted in Fig.\ref{fig:Fig12}(a) and (b), respectively. In Ta-Ta, at $\Gamma$, the transmission coefficient is only $1.98\times10^{-4}$ which is entirely contributed by $\emph{d}{_{z^2}}$ orbitals. In W-W, where both $\emph{d}{_{z^2}}$ and $\emph{d}{_{xz,yz}}$ orbitals populate at $k_\parallel=(0,0)$, a moderate coefficient of $9.4\times10^{-3}$ is obtained. Almost negligible transmission of $3.9\times10^{-14}$ is found close to Fermi energy in Mo-Mo because the occupancy of $\emph{d}{_{xz,yz}}$ orbital is the largest at the Fermi energy. The orbital characters compatible to propagate at normal incidence i.e. at $\Gamma$-point are the $\emph{s}$, $\emph{p}{_z}$ and $\emph{d}{_{z^2}}$ orbitals that build $\Delta_1$ Bloch states in crystals with square symmetry \cite{butler2008tunneling}. Noteworthy, in Mo-Mo MTJ, $\emph{d}{_{xz,yz}}$ is still a prime out-of-plane orbital character compatible with the $\Delta_5$ wave symmetries. In principle, the contribution of $\Delta_5$ symmetries to the transmission is out-of-normal to incidence i.e. away from $k_\parallel=(0,0)$ which is not depicted in Fig.\ref{fig:Fig4}.  Besides, these symmetries certainly show a strong influence on the transmission in P$^{\downarrow \downarrow}$ and AP$^{\uparrow \downarrow/\downarrow \uparrow}$ channels. As a result, a relative increment is observed in G$_P^{\downarrow\downarrow}$ and G$_{AP}^{\uparrow \downarrow/\downarrow \uparrow}$ in the Mo-Mo MTJ, see Fig.\ref{fig:Fig2}(a). Moreover, a sharp transmission peak comparatively an order of magnitude greater than W-W is observed in IrMn-IrMn MTJ which is majorly contributed by $\emph{d}{_{z^2}}$ orbitals. This increase is due to the optimal lattice-mismatch-induced strong interfacial coupling which creates an easy conducting channel for electrons close to the normal incidence. This conduction channel is contributed by the out-of-plane $\emph{d}{_{z^2}}$ character of IrMn as seen in Fig.\ref{fig:Fig12}(a). 
 
\section{Orbital Reconstruction in Mo-Mo MTJ}
\begin{figure}[h!]
\centering
\includegraphics[width = 0.9\columnwidth]{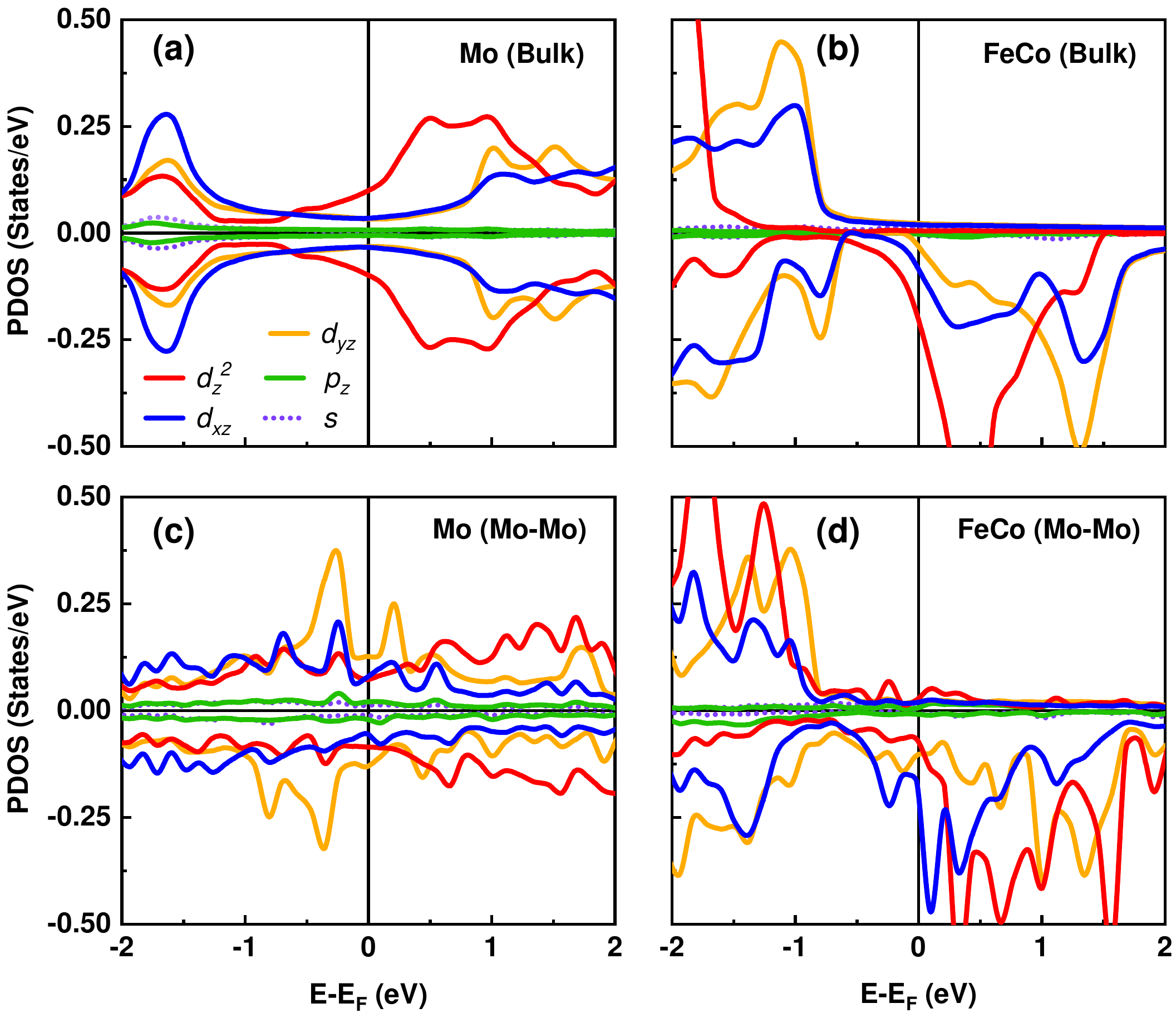}
\caption{Orbital-reconstruction. Density-of-states projected on the out-of-plane orbitals for (a) bulk Mo and (b) bulk FeCo. Also, for (c) 5MLs of Mo-capping (d) 5MLs of FeCo-electrode, in Mo-Mo MTJ. The Fermi level is set to 0 eV.}
\label{fig:Fig13}
\end{figure}

The absence of $\emph{d}{_{z^2}}$ ($\Delta_1$ states) at $k_{\parallel}=(0,0)$ in Fig.\ref{fig:Fig5}(g) is anticipated to have caused by the change in electronic structure at the Mo/FeCo interface or in the Mo and/or FeCo layers, itself.  As a results, an imbalance in orbital-construction and hence, in the orbital hybridization occurs.  Besides, it is already known that Mo alters the (001) crystallography of the adjacent bcc FeCo-electrode \cite{almasi2016effect}. To further confirm the impact of Mo capping layer on the electronic structure, we examine orbital-construction within the bulk Mo (Fig.\ref{fig:Fig13}(a)) and the bulk FeCo (Fig.\ref{fig:Fig13}(b)) where, clearly, $\emph{d}{_{z^2}}$ ($\Delta_1$-states) is found to dominate at the Fermi level. In contrast, the orbital-reconstruction plotted in Fig.\ref{fig:Fig13}(c) and (d), as a consequence of stacking Mo/FeCo together in Mo-Mo MTJ, indicate the precedence of $\emph{d}{_{xz,yz}}$ ($\Delta_5$-states) over $\emph{d}{_{z^2}}$ ($\Delta_1$-states) at the Fermi level.

\section{Scattering States for W-W MTJ}
\begin{figure}[h!]
\centering
\includegraphics[width = 1.0\columnwidth]{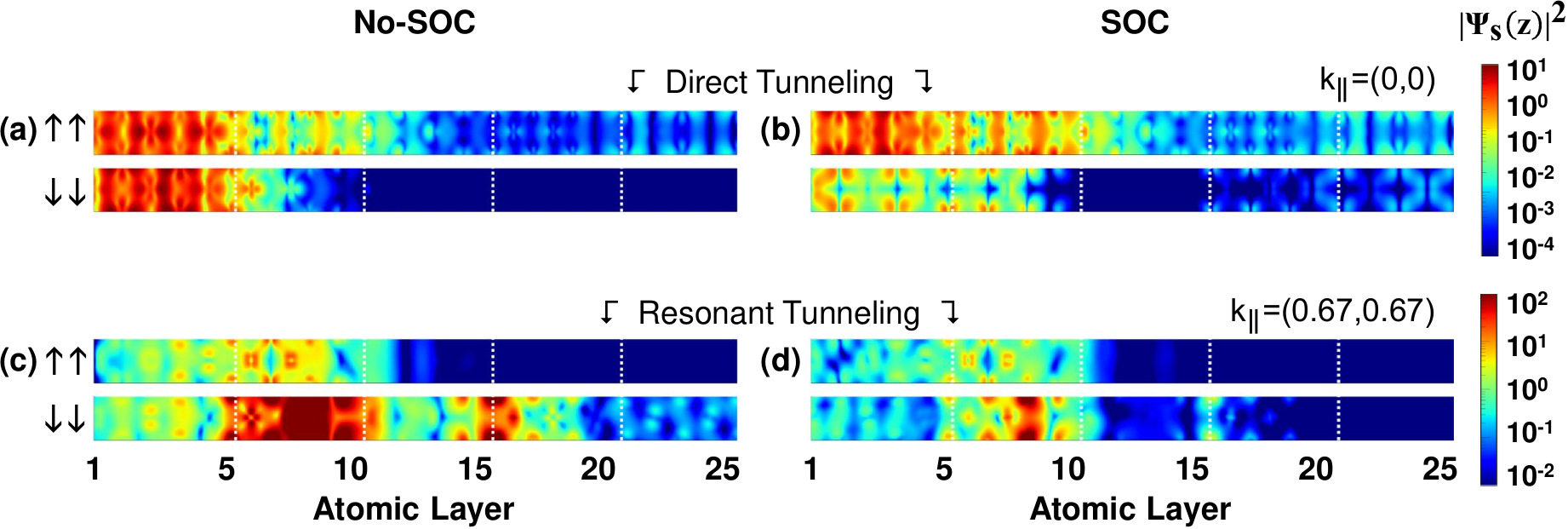}
\caption{Real-space distribution of squared modulus of scattering states ${|\Psi_s(z)|^2}$ in parallel configuration across the W-W MTJ, under equilibrium, at the Fermi energy. The scattering states are calculated at different $k_\parallel$  points: (a),(b) $k_\parallel=(0,0)$ and (c),(d) $k_\parallel=(0.67,0.67)$, without and with SOC. The values of $k_\parallel$  are given in the units of $2\pi/a$. In (a) P$^{\uparrow\uparrow}$ indicates 'direct tunneling' mechanism at $\Gamma$-point, while in (b) P$^{\downarrow\downarrow}$ indicates 'Resonant tunneling' governed by IRS. The vertical dashed lines roughly mark the interfaces. The color bar represents logarithmic scaled $|\Psi_s(z)|$  when the spin-wave travels along the direction of transport (z-direction).}
\label{fig:Fig14}
\end{figure}

The scattering wave at $k_{\parallel}$=0,0 in  P$^{\uparrow\uparrow}$ channel in Fig.(\ref{fig:Fig14}(a) and (b) shows mixed composition of \emph{d}$_{z^2}$ and $\emph{d}{_{xz,yz}}$ orbital characters of W (Fe, Co) and \emph{p}$_z$ character of O in the MgO-barrier. Similar to IrMn-IrMn MTJ, the scattering wave in P$^{\uparrow\uparrow}$ of Fig.(\ref{fig:Fig14}(b) shows stronger transmission after switching on SOC than in Fig.(\ref{fig:Fig14}(a). This is, again, attributed to an increase in band (orbital) levels due to the SOC-induced lifting of spin degeneracy. Next, an obvious change in the dominating orbital character of the scattering wave in P$^{\downarrow\downarrow}$ channels is seen in Fig.(\ref{fig:Fig14}(a) and (b). When SOC is switched on,  P$^{\downarrow\downarrow}$ scattering wave is dominated by the $\emph{d}{_{xz,yz}}$ orbital characters. This scattering wave tunnels to the other side of the MgO-barrier, however, with lower magnitude than in IrMn-IrMn MTJ. The SOC-induced orbital-mixing forms this additional conduction channel and encourages  $\emph{d}{_{xz,yz}}$ orbitals to transmit at normal incidence. Importantly, at $k_{\parallel}=(k_x,k_y)$, the strength of localized IRS P$^{\downarrow\downarrow}$ channels in Fig.\ref{fig:Fig14}(d) is found to have considerably reduced as compared to P$^{\downarrow\downarrow}$ in Fig.\ref{fig:Fig14}(c). In fact, in addition, there is no significant wave tunneling that occurs in P$^{\uparrow\uparrow}$ of Fig.(\ref{fig:Fig14}(d) as in P$^{\uparrow\uparrow}$ of Fig.(\ref{fig:Fig8}(d) in IrMn-IrMn MTJ.

\bibliography{citations}

\end{document}